\documentclass[a4paper, amsfonts, amssymb, amsmath, reprint, showkeys, nofootinbib, twoside,notitlepage,onecolumn]{revtex4-2}

\def\apj{{ApJ}}                 
\def\apjs{{ApJS}}               
\def\aap{{A\&A}}                

\def\mnras{{MNRAS}}             
\def\prd{{Phys.~Rev.~D}}        
\def\jqsrt{{J.~Quant.~Spec.~Radiat.~Transf.}}

\usepackage{amsmath,amstext}
\usepackage[T1]{fontenc}
\usepackage{amssymb}
\input{epsf}
\usepackage{graphicx}
\usepackage{ae,aecompl}

\usepackage{hyperref}
\usepackage{amsmath}
\usepackage{amssymb}
\usepackage{mathtools}
\usepackage{bm}
\usepackage{cleveref}
\usepackage{tensor}
\usepackage{braket}
\usepackage{enumitem}
\usepackage{mhchem}
\usepackage{cancel}
\usepackage{amsthm}
\usepackage{upgreek}
\usepackage{mathrsfs}
\usepackage{slashed}

\usepackage{color}

\newcommand{\spc}{\quad \quad \quad}

\newcommand{\+}{{\textcolor{white}{+}}}

\def\be{\begin{equation}}
\def\ee{\end{equation}}
\def\beq{\begin{eqnarray}}
\def\eeq{\end{eqnarray}}

\theoremstyle{definition}

\theoremstyle{theorem}

\theoremstyle{corollary}

\begin{document}

\title{Non-Newtonian corrections to radiative viscosity: Israel-Stewart theory as a viscosity limiter}
\author{ L.~Gavassino}
\affiliation{
Department of Mathematics, Vanderbilt University, Nashville, TN, USA
}

\begin{abstract}
Radiation is a universal friction-increasing agent. When two fluid layers are in relative motion, the inevitable exchange of radiation between such layers gives rise to an effective force, which tries to prevent the layers from sliding. This friction is often modeled as a Navier-Stokes shear viscosity. However, non-Newtonian corrections are expected to appear at distances of about one optical depth from the layers' interface. Such corrections prevent the viscous stress from becoming too large. Here, we set the foundations of a rigorous theory for these corrections, valid along incompressible flows. We show that, in the linear regime, the infinite Chapman-Enskog series can be computed analytically, leading to universal formulas for all transport coefficients, which apply to any fluid, with any composition, with radiation of any type (also neutrinos), and with nearly any type of radiative process. We then show that, with an appropriate shear-heat coupling coefficient, Israel-Stewart theory can correctly describe most non-Newtonian features of radiative shear stresses.
\end{abstract} 

\maketitle
\vspace{-0.5cm}

\begin{figure}
    \centering
    \includegraphics[width=0.44\linewidth]{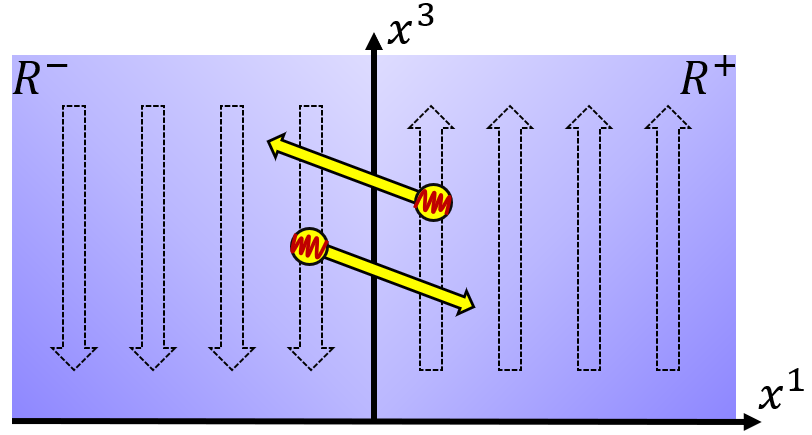}
\caption{Microscopic origin of radiative shear viscosity. Consider a fluid that moves along the $x^3$ axis, with some velocity $v>0$ in the region $R^+=\{x^1{>}0\}$, and opposite velocity $-v$ in $R^-=\{x^1{<}0\}$. On average, radiation particles (yellow) emitted in $R^-$ carry negative momentum in direction $x^3$, since their source is negatively boosted. Hence, if they are reabsorbed inside $R^+$, they will impress (on average) a negative kick on the fluid. Similarly, radiation emitted from $R^+$ impresses a positive kick when absorbed in $R^-$. The result is a frictional force between the fluid layers, a.k.a. a shear viscosity. }
    \label{fig:friction}
\end{figure}

\noindent\textbf{\textit{Introduction -}} Our ability to model high-mass stars, optically thick accretion disks, supernovae, and neutron-star mergers heavily depends on our understanding of how matter and thermal radiation (photons or neutrinos) interact at macroscopic (i.e. fluid-dynamic) scales, see e.g. \cite{Bruenn1985,Rampp2002,Rampp:2002bq,Farris2008,Chiavassa2011,Fragile:2014bfa,Mirizzi:2015eza,Radice:2016dwd,Murchikova:2017zsy,Fragile:2018xee,Lancova:2019ldt,Perego:2019adq,Anninos2020,Fukushima2021,Foucart:2022bth,RadiceNew2022,Wang:2022voe}. This problem is so important for astrophysical modeling that there is a dedicated branch of fluid mechanics, called ``radiation hydrodynamics'', whose central goal is to develop effective descriptions of the radiation field, which may be implemented in numerical simulations \cite{Pomraning1973,mihalas_book,CastorRadiationBook_2004}.
Most aspects of the radiation-hydrodynamic problem (e.g. radiative heat conduction and bulk viscosity \cite{NovikovThorne1973,weinbergQFT_1995,UdeyIsrael1982,Minerbo1978,Levermore1984,GavassinoRadiazione}) are at this point well-understood, and current numerical schemes, especially the M1 closure \cite{Minerbo1978,Levermore1984,ShibataRadiation2011k,Sadowski2013,RadiceNew2022,GavassinoRadiazione}, capture the essential physics quite reliably.
However, to date, radiation hydrodynamics has failed to provide a systematic and rigorous theory for radiative \textit{shear} viscosity, whose foundations we aim to set here.

\noindent\textit{\textbf{State of the art -}} Let us first review what is currently known. The microscopic picture is very simple \cite[\S 97]{mihalas_book}: Two fluid layers in relative motion exchange radiation, and in doing so redistribute their momentum, thereby reducing their relative speed, see figure \ref{fig:friction}\footnote{Throughout the article, we work in Minkowski spacetime, with metric signature $(-,+,+,+)$, and we adopt natural units, such that $c=\hbar=k_B=1$. Greek indices run from 0 to 3 (with $x^0=t$), while Latin indices run from 1 to 3.}. At the fluid-dynamic scale, this has proven surprisingly hard to model accurately. It was suggested in \cite{Thomas1930,Misner1968,Weinberg1971} that radiation viscosity should follow ordinary Navier-Stokes theory. However, this approximation is valid only when the radiation mean free path is infinitesimal compared to the lengthscale of the hydrodynamic gradients, which is not the case in many real systems. Indeed, the authors of \cite[\S 6.2]{ShibataRadiation2011k} noted that sometimes the radiative Navier-Stokes stress ends up being larger than the radiative pressure, and introduced (without derivation) a ``viscosity limiter'', in the same spirit as flux limiters are adopted for radiative heat conduction \cite[\S 11.5]{CastorRadiationBook_2004}. In \cite{UdeyIsrael1982}, it was argued that radiative shear stresses should instead be described by Israel-Stewart theory \cite{Israel_Stewart_1979,Hishcock1983,WagnerGavassino2023jgq}, but there currently is no evidence for such claim (besides stability arguments \cite{Hiscock_Insatibility_first_order,PuKoide2009fj,GavassinoLyapunov_2020,GavassinoSuperluminal2021}). Thus, no consensus has been reached.

Given these difficulties, shear contributions to the radiative stress tensor are set to zero in current simulations \cite{Farris2008,Sadowski2013,OConnor:2014sgn,Pan:2018vkx,Bloch2021,Bloch2022,RadiceNew2022,Schianchi:2023uky} (except, of course, in full Boltzmann solvers \cite{Sumiyoshi:2012za,Nagakura:2017mnp,Bhattacharyya:2022bzf}). Hence, simulated shear waves cannot decay, and the layers in figure \ref{fig:friction} are doomed to slide forever, see Supplementary Material (section II) for the proof.

\noindent\textbf{\textit{Problem statement -}} To make objective (i.e. quantifiable) progress, we need a mathematically well-defined problem. We propose the following setup. Consider again the flow shown in figure \ref{fig:friction}, and use it as an initial condition for the hydrodynamic evolution, namely $u^3(0,x^1)=v \,  \text{sgn}(x^1)$ (assuming $v$ small). If the viscous stress had a Navier-Stokes form, as suggested in \cite{Thomas1930,Misner1968,Weinberg1971}, then the flow velocity component $u^3$ would obey the diffusion equation $W\partial_t u_3\,{=}\,\eta^{(1)}\partial_1^2 u_3$, where $\eta^{(1)}$ is the shear viscosity coefficient, and $W$ the inertia density of the fluid. The evolution would then be
\begin{equation}\label{noupdate}
u^3(t,x^1)
=
v \,
\text{Erf}\bigg( \dfrac{x^1}{\sqrt{4\eta^{(1)} t/W}}\bigg)\, .
\end{equation}
On the other hand, if we solved the radiative transport equation exactly, we would find deviations from \eqref{noupdate} over distances of one radiation mean free path away from $x^1{=}0$. To be useful, a non-Newtonian modification of Navier-Stokes should correctly resolve the flow's structure at those scales. Due to the analogy with rheometric experiments in material engineering \cite{Malkin_book,Steffe_book}, we may interpret this as the fundamental problem in ``radiative rheology''.

In this Letter, we show that, if $v$ is small and $W$ is large (i.e. in matter-dominated fluids that are sliding slowly), the above problem can be approached with fully analytical techniques, and the following remarkable facts hold:
\begin{itemize}
\item[(i)] The linearized radiative transfer equation admits an explicit solution in Fourier space, for flows as in figure \ref{fig:friction}. Hence, the microscopic corrections to \eqref{noupdate} can be computed \textit{exactly}, leading to formulas that are valid for fluids with any composition, radiation of any kind, and almost any type of radiation-matter interaction process.
\item[(ii)] The shear stress $\Pi_{13}$ computed from point (i) can be expanded as a Chapman-Enskog gradient series \cite{McLennan1965,Dudynski1989,Struchtrup,GavassinoChapmanEnskog2024xwf},
\begin{equation}\label{pi}
 \Pi_{13}=-\sum_{a=1}^{+\infty}\eta^{(2a-1)} \partial^{2a-1}_1  u_3 \, ,
\end{equation}
and there is an exact (non-recursive) formula for all transport coefficients $\eta^{(2a-1)}$, see equation \eqref{etini}. In Fourier space, the radius of convergence of the above series coincides with the magnitude of the non-hydrodynamic gap.
\item[(iii)] Most features of the exact solution in point (i) are reproduced remarkably well by the ``viscosity-limited'' model
\begin{equation}\label{nonNewt}
\bigg(1-\dfrac{\eta^{(1)}}{\eta^{(-1)}}  \partial^2_1\bigg)\Pi_{13}(x^1)=\Pi^{\text{NS}}_{13}(x^1) \quad \quad \Longleftrightarrow \quad \quad \Pi_{13}(x^1)= \int_{-\infty}^{+\infty} \dfrac{e^{-|\xi|}}{2} \,  \Pi^{\text{NS}}_{13}\bigg(x^1{+} \xi \sqrt{\frac{\eta^{(1)}}{\eta^{(-1)}}} \bigg)  \, d\xi \, ,
\end{equation}
where $\eta^{(-1)}$ is also given by \eqref{etini}, and $\Pi^{\text{NS}}_{13}=-\eta^{(1)}\partial_1 u_3$ is the Navier-Stokes shear stress.
\item[(iv)] The non-Newtonian model \eqref{nonNewt} is the large-$W$ limit of the transversal gapless mode of an Israel-Stewart theory that contains both the shear stress $\Pi_{\mu \nu}$ and the heat flux $q_\mu$, see \cite[Eq.s (77), (78)]{GavassinoNonHydro2022}, with a shear-heat coupling coefficient given by  $\alpha_1{=}\pm (T\kappa_q \, \eta^{(-1)})^{-1/2}$, where $\kappa_q$ is the (radiative \cite{NovikovThorne1973}) heat conductivity.
\end{itemize}
The derivation of (i,ii,iii,iv) is provided below.

\noindent\textbf{\textit{\textbf{Geometric considerations -}}} Following \cite{Thomas1930,Weinberg1971,UdeyIsrael1982}, we consider an ideal (possibly multiconstituent) fluid with negligibly short mean free path, which interacts with a gaseous mixture of radiation quanta (photons and/or neutrinos) with finite mean free path (whose size may be comparable to the geometry of the flow). The fluid is assumed to be in local kinetic equilibrium, but chemical reactions are allowed. Hence, the local state of the fluid is characterized by a single flow velocity field $u^\mu(x^\alpha)$, and by a list of thermodynamic scalars $\Lambda(x^\alpha)=\{\text{``temperature''},\text{``densities''}...\}$. The radiation gas, instead, may evolve away from local kinetic equilibrium, and is modeled using kinetic theory. We will be therefore tracking the invariant distributions $f^A_{\textbf{p}}(x^\alpha)$, which count how many radiation particles of type $A$ occupy a one-body state located at $x^\alpha$ and with momentum $\textbf{p}$. Hence, the degrees of freedom of the composite ``fluid+radiation'' system at a given event $x^\alpha$ are
\begin{equation}
\Psi(x^\alpha)=\big[\Lambda,\, u^1, \, u^2,\, u^3, \, f^A_{p^1,p^2,p^3}\big]\, .
\end{equation}

Fix some global equilibrium state $\Psi{=}\text{``constant''}$, with zero velocity, and let $\delta \Psi(t,x^1)$ be a linear plane-wave perturbation away from $\Psi$. Call $\delta\Tilde{\Psi}(t,x^1)$ the result of applying to $\delta \Psi(t,x^1)$ a 180-degree rotation around the $x^1$-axis. Then, for a given $(t,x^1)$, the values of $\delta \Psi$ and $\delta \Tilde{\Psi}$ are related as follows (we assume unpolarised radiation):
\begin{subequations}
\begin{eqnarray}
\delta\Psi(t,x^1)={}& \big[\delta\Lambda,\, \+ \delta u^1, \, \+ \delta u^2,\, \+ \delta u^3, \, \+ \delta f^A_{p^1,p^2,p^3}\big] \, ,\\
\delta\Tilde{\Psi}(t,x^1)={}& \big[\delta\Lambda,\, \+ \delta u^1, \, -\delta u^2,\, -\delta u^3, \, \+ \delta  f^A_{p^1,-p^2,-p^3}\big] \, .
\end{eqnarray}
\end{subequations}
Now, if $\delta \Psi(t,x^1)$ is a solution of some linearised equations of motion, then, since the equilibrium state is isotropic, also $\delta \Tilde{\Psi}(t,x^1)$ solves the same equations. Indeed, by linearity,  also the following linear combinations are solutions:
\vspace{-0.1cm}
\begin{subequations}\label{longitudinal}
\begin{eqnarray}
\label{lingit}\delta\Psi_L(t,x^1)=\dfrac{\delta \Psi(t,x^1){+}\delta \Tilde{\Psi}(t,x^1)}{2}={}& \bigg[\delta\Lambda,\,  \delta u^1, \, 0,\, 0, \, \dfrac{\delta f^A_{p^1,p^2,p^3}+ \delta f^A_{p^1,-p^2,-p^3}}{2}\bigg] \, ,\\
\label{shir}\delta\Psi_T(t,x^1)=\dfrac{\delta \Psi(t,x^1){-}\delta \Tilde{\Psi}(t,x^1)}{2}={}& \bigg[0,\,  0, \, \delta u^2 ,\, \delta u^3, \, \dfrac{\delta f^A_{p^1,p^2,p^3}- \delta f^A_{p^1,-p^2,-p^3}}{2}\bigg] \, ,  
\end{eqnarray}
\end{subequations}
where $\delta \Psi_L$ is the \textit{longitudinal flow} (which is compressible) and $\delta \Psi_T$ the \textit{transversal flow} (which is incompressible). It is evident that the flow in figure \ref{fig:friction} is transversal, which is convenient, since $\delta \Lambda$ vanishes on transversal flows, and this considerably simplifies the rheological problem.

\noindent \textbf{\textit{Transversal evolution -}} Following figure \ref{fig:friction}, let us assume that $\delta u^2=0$. Then, we see from equation \eqref{shir} that the only relevant degrees of freedom of a transversal flow are $\{ \delta u^3, \delta f^A_\textbf{p}\}$, the latter being subject to the constraint $\delta f^A_{p^1,p^2,p^3}{=}-\delta f^A_{p^1,-p^2,-p^3}$. In most scenarios, these variables evolve according to the following generic equations of motion, valid for photons, neutrinos, and other hypothetical radiation particles (e.g. gravitons \cite{Weinberg1971}):
\vspace{-0.1cm}
\begin{subequations}
		\label{eom}
		\begin{eqnarray}
&\label{conservation} W \partial_t \delta u^3 +{\displaystyle\sum_{A} 
g_A }{\displaystyle\int_{\mathbb{R}^3}} \dfrac{d^3 p}{(2\pi)^3} p^3 \, \bigg(\partial_t{+}\dfrac{p^1}{p^0} \partial_1\bigg) \delta f^A_{\textbf{p}}=0 \, , \\
&\label{boltzmann} \bigg(\partial_t {+}\dfrac{p^1}{p^0} \partial_1 \bigg)\delta f^A_{\textbf{p}} =\mathcal{S}^A(p^0) \, p^3 \delta u^3  -{\displaystyle \sum_B\mathcal{M}\indices{^A _B}(p^0) }\,  \delta f^B_{\textbf{p}} \, ,
\end{eqnarray}
\end{subequations}
where $g_A$ is the spin degeneracy of the component $A$. The coefficients $\{\mathcal{S}^A , \mathcal{M}\indices{^A _B}\}$ are two interaction terms, both of which are functions of the energy $p^0=\sqrt{p^j p_j}$.

Let us unpack the meaning of \eqref{eom}. Equation \eqref{conservation} is the conservation of momentum ($\partial_\mu \delta T^{\mu 3}_{\text{fluid}}{+}\partial_\mu \delta T^{\mu 3}_{\text{rad}}{=}0$) for the composite ``fluid+radiation'' system, where the fluid stress-energy tensor $T^{\mu \nu}_{\text{fluid}}$ has the usual ideal-fluid form, while the radiation stress-energy tensor $T^{\mu \nu}_{\text{rad}}$ is that of an ultrarelativistic gas mixture (modeled using kinetic theory). Equation \eqref{boltzmann} is the radiative Boltzmann equation. The form of the collision term on the right-hand side is almost universal. Indeed, we demonstrate in the Supplementary Material that \eqref{boltzmann} can account for absorption, emission, isotropic scattering, pair processes, and flavor oscillations in an \textit{exact} manner, and we compute the relative contributions to $\{\mathcal{S}^A , \mathcal{M}\indices{^A _B}\}$. The key observation that leads to the universality of \eqref{boltzmann} is the following. According to equation \eqref{shir}, the perturbation $\delta f^A_{\textbf{p}}$ of a shear wave is odd under the rotation $(p^1,p^2,p^3)\rightarrow (p^1,-p^2,-p^3)$, so that $\int F(p'^0) \, \delta f^A_{\textbf{p}'} d^3 p' \, {=} \,0$
for any function $F(p^0)$. This causes isotropic Boltzmann integrals involving $\delta f^A_{\textbf{p}'}$ to vanish, leaving only the two terms in \eqref{boltzmann}. On a longitudinal wave, this simplification would not occur, again leading to non-universal behaviors\footnote{The non-universal behaviour of longitudinal waves is due to the fact that compressible flows are sensitive to bulk viscosity, whose phenomenology is known to be very rich and complex \cite[\S 81]{landau6}, especially in relativity \cite{BulkGavassino,Hydro+2018,GavassinoFarFromBulk2023,GavassinoBurgers2023}. Luckily, the M1 closure scheme currently used in numerical simulations is known to reproduce radiative bulk viscous effects correctly \cite{GavassinoRadiazione}.}.

\noindent\textbf{\textit{Shear waves -}}
We solve \eqref{eom} for shear waves, i.e. sinusoidal transversal plane waves, with a spacetime dependence of the form $e^{-i(kx^1-\omega t)}$. Then, equation \eqref{boltzmann} gives\footnote{Note that \eqref{pipertino} has the form $\delta f^A(\textbf{p})=K^A(p^0,p^1) p^3 $, which is indeed odd for 180$^o$ rotations around the $x^1$ axis, consistently with \eqref{shir}.}
\begin{equation}\label{pipertino}
\delta f^A_\textbf{p}=\sum\nolimits_{B} \big[(\mathcal{M}-i\omega+ikp^1/p^0)^{-1} \big]\indices{^A _B} \,  \mathcal{S}^B p^3 \delta u^3 \, ,
\end{equation}
which can be plugged into \eqref{conservation}. Let us assume that $\mathcal{M}$ is diagonalizable and invertible, namely $\mathcal{M}=\sum_n \tau_n^{-1} \mathcal{P}_n$, with inverse eigenvalues $\tau_n \in \mathbb{C}$ (we interpret them as ``mean free paths''$\, +\, i \, \times \,$``oscillation times''), and eigenprojectors $\mathcal{P}_n$, such that $\mathcal{P}_m \mathcal{P}_n=\delta_{mn}\mathcal{P}_n$ and $\sum_n \mathcal{P}_n=1$. Then, expressing the momentum integral in \eqref{conservation} in spherical coordinates ($d^3p=E^2 dE d^2\Omega$), we can solve the angular integral analytically, and we obtain the following \textit{exact} implicit formula:
\begin{equation}\label{omega}
\omega +\dfrac{i}{W} \sum_n \int_0^{\infty}  \bigg[\dfrac{2}{3}+\dfrac{1{-}i\omega \tau_n}{(k\tau_n)^2}-\bigg(1+ \dfrac{(1{-}i\omega \tau_n)^2}{(k\tau_n)^2} \bigg) \dfrac{1}{k\tau_n} \arctan\bigg( \dfrac{k\tau_n}{1{-}i\omega \tau_n}\bigg)\bigg] \rho_n dE=0 \, ,
\end{equation}
where we stress that $\tau_n$ and $\rho_n$ are (possibly complex) functions of the energy $E$, the latter being given by
\vspace{-0.1cm}
\begin{eqnarray}\label{rhon}
\rho_n(E) =\dfrac{E^4}{4\pi^2} \sum_{A,B} g_A [\mathcal{P}_n(E)]\indices{^A _B} \mathcal{S}^B(E) \, .
\end{eqnarray}
Note the generality of this result: The frequency $\omega$ of shear waves has the universal (implicit) functional dependence \eqref{omega} on the wavenumber $k$, regardless of the matter composition, the type of radiation, and the details of the interaction. All that information is stored in the functions $\{\tau_n(E),\rho_n(E) \}$, which quantify respectively the lifetime of radiation and the emissivity of the fluid. In the Supplementary Material, we plot \eqref{omega} in the particular case of a grey material.
 \newpage


\noindent\textbf{\textit{Effective viscous theory -}} Let us now recall our main goal: We aim to describe radiative effects as viscous corrections to the motion of the fluid. Hence, we are mostly interested in matter-dominated systems, where the inertia of radiation is small compared to the inertia of the fluid (i.e. $W\rightarrow \infty$). In this limit, if $|\tau_n(E)|$ is bounded above, we have that\footnote{\label{bigproofs}Proof: Fix $k\in \mathbb{R} {\setminus} \{0\}$, and regard ``$-i\times $\eqref{omega}'' as an implicit function $F(\lambda, \Gamma)\,{=}\, 0$, with $\lambda \, {=} \, W^{-1}$ and $\Gamma \, {=} \, {-}i\omega$ (note that $F:\mathbb{R}^2 \rightarrow \mathbb{R}$). We are interested in the solution to \eqref{omega} in a neighborhood of the point $(\lambda,\Gamma)=(0,0)$, corresponding to a fluid component with infinite inertia. We note that the following facts hold: (a) $F(0,0)=0$, (b) $\partial_\Gamma F(0,0)\neq 0$, and (c) $F$ is smooth around $(0,0)$. Then, the implicit function theorem applies, and there exists locally a unique smooth function $\Gamma(\lambda)$ such that $\Gamma(0)=0$, and $F(\lambda,\Gamma(\lambda))=0$. This function can be expanded to first order in $\lambda$, giving $\Gamma(\lambda)= \Gamma'(0)\lambda+\mathcal{O}(\lambda^2)$, where $\Gamma'(0)=-\partial_\lambda F(0,0)/\partial_\Gamma F(0,0)$. The result is \eqref{omegaalphato}.} 
\begin{equation}\label{omegaalphato}
\omega =-\dfrac{i}{W} \sum_n \int_0^{\infty}  \bigg[\dfrac{2}{3}+\dfrac{1}{(k\tau_n)^2}-\bigg(1{+} \dfrac{1}{(k\tau_n)^2} \bigg) \dfrac{\arctan (k\tau_n)}{k\tau_n} \bigg] \rho_n dE \, .
\end{equation}
The function $\omega(k)$ is now explicit. We can formally expand it in powers of $k$, as long as $k<1/\sup\{|\tau_n(E)|\}$, giving
\begin{equation}\label{serione}
\omega=\dfrac{i}{W} \sum_{a=1}^{\infty} \dfrac{2 (ik)^{2a}}{(1{+}2a)(3{+}2a)} \int_0^{\infty}  \sum_n \rho_n \tau_n^{2a} dE \, .
\end{equation}
This can be reinterpreted as the dispersion relation of a viscous fluid, governed by an effective equation of motion of the form $W\partial_t  u_3 +\partial_1  \Pi_{13}=0$, with an effective viscous shear stress $  \Pi_{13}$ given by the gradient expansion \eqref{pi}.
Comparing \eqref{serione} with \eqref{pi}, we obtain an exact formula for all the infinite transport coefficients of the shear channel:
\begin{equation}\label{etini}
\eta^{(2a-1)}=\dfrac{2}{(1{+}2a)(3{+}2a)} \int_0^\infty \sum_n \rho_n \tau_n^{2a} dE \, .
\end{equation}
This is a very rare occurrence: A system with fully realistic interactions where the Chapman-Enskog expansion can be computed analytically up to infinite order. In grey media where only absorption, emission, and isoenergetic scattering take place, our (exact) shear viscosity coefficient $\eta^{(1)}$ coincides with that computed in \cite{Thomas1930,Misner1968,Weinberg1971}, see also \cite{GavassinoRadiazioDispersion2024cqw}.

\noindent\textbf{\textit{Non-Newtonian model -}} Since the gradient expansion \eqref{pi} has a finite radius of convergence, we cannot use it to model radiation whose mean free path $\tau= \sup\{|\tau_n(E)|\}$ is too long. Thus, the natural question is: Is there a simple partial differential equation with a \textit{finite} number of derivatives that well approximates \eqref{omegaalphato} both at small and at large $k\tau$? To agree with \eqref{omegaalphato}, the dispersion relation $\omega(k)$ arising from such an equation should have the following limits:
\begin{equation}\label{request}
\omega(k) \approx - \dfrac{i}{W} \times
\begin{cases}
\eta^{(1)} k^2 & \text{if }k\tau \rightarrow 0 \, , \\
\eta^{(-1)}  & \text{if }k\tau \rightarrow \infty \, ,
\end{cases}
\end{equation}
where $\eta^{(-1)}$ is defined by formally setting $a=0$ in \eqref{etini}. Since $\eta^{(1)}\sim \eta^{(-1)}\tau^2$, the most natural ``interpolation'' is
\begin{equation}\label{limitsforti}
\omega(k) = -\dfrac{i}{W} \, \dfrac{\eta^{(-1)}\eta^{(1)}k^2}{\eta^{(-1)}+\eta^{(1)}k^2}  \equiv \dfrac{k\Pi_{13}}{W u^3} \, ,
\end{equation}
which works reasonably well also at intermediate $k \tau$, albeit less accurately than \eqref{pi}. The linear partial differential equation associated with \eqref{limitsforti} is
\begin{equation}\label{themodelone}
W\bigg(1-\dfrac{\eta^{(1)}}{\eta^{(-1)}} \partial^2_1 \bigg)\partial_t  u_3  =\eta^{(1)} \partial^2_1  u_3 \, .
\end{equation}
Furthermore, if we take the inverse Fourier transform of the second equal sign in \eqref{limitsforti}, we recover \eqref{nonNewt}.

Let us confirm that this simple model constitutes an objective improvement to the Navier-Stokes theory, by explicitly verifying that it captures the evolution of the state in figure \ref{fig:friction} more accurately than equation \eqref{noupdate}. To this end, let us first note that, if the initial data is $u^3(0,x^1)=v \,  \text{sgn}(x^1)$, the solutions to respectively the Navier-Stokes equation, the model \eqref{themodelone}, and the full system \eqref{eom} can all be written as Fourier integrals of the form
\begin{equation}\label{fourore}
 u^3 (t,x^1)= v \int_0^{\infty} \dfrac{2\sin(kx^1)}{\pi k} e^{-i\omega(k)t} dk \, ,
\end{equation}
each with its corresponding dispersion relation. Since all these models agree at small $k$ by construction, they all share the same late-time behavior, by the relaxation effect \cite{LindblomRelaxation1996}. The real question is what happens at early times near $x^1 = 0$. Here, the trick is to note that, if we make the
change of variable $d\xi =x^1 dk$ in \eqref{fourore}, and send $x^1$ to 0, the integral gives
$u^3(t,0^\pm)=\pm ve^{-i\omega(\infty)t}$. From equation \eqref{request}, we see that the full radiative transport solution has a jump discontinuity of size $2v \, e^{-\eta^{(-1)}t/W}$, which is correctly reproduced by \eqref{themodelone}, while is absent in Navier-Stokes. This means that, while \eqref{themodelone} is aware that the fluid layers $\{x^1\,{=}\,0^+\}$ and $\{x^1\,{=}\,0^-\}$ slide against each other over a relative distance $2vW/\eta^{(-1)}$ before being stopped by friction, the Navier-Stokes theory predicts that they stop immediately. The reason for this shortcoming is that Navier-Stokes theory is not ``viscosity-limited'', and thus generates infinite stresses near discontinuities. By contrast, radiative kinetic theory is viscosity-limited, as shown in equation \eqref{request}.

\newpage

\begin{figure}
    \centering
\includegraphics[width=0.49\linewidth]{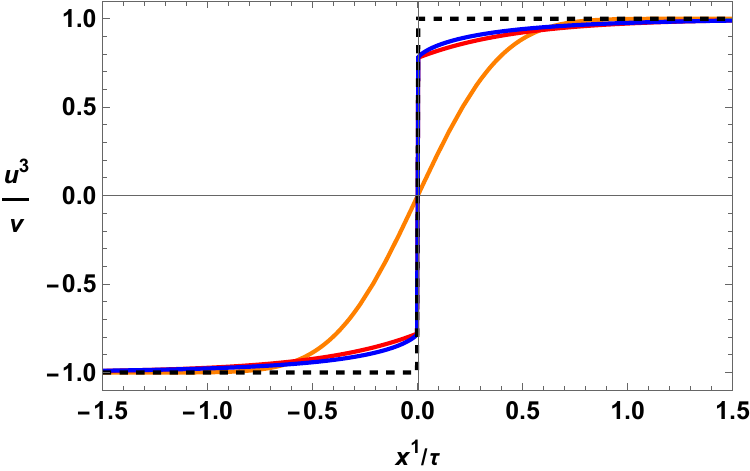}
\includegraphics[width=0.49\linewidth]{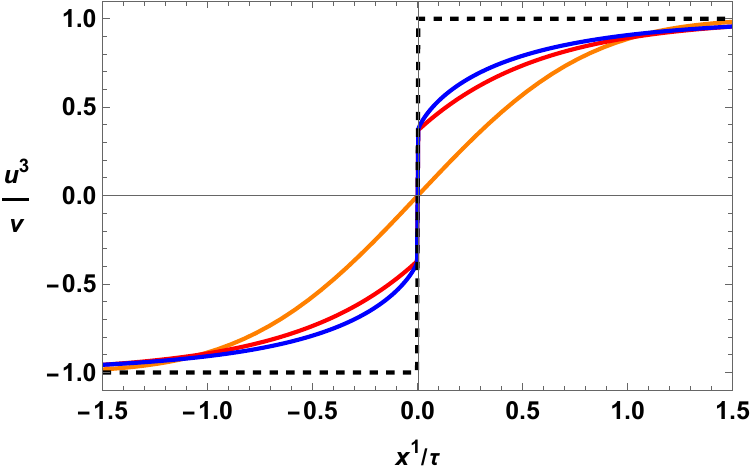}
\includegraphics[width=0.49\linewidth]{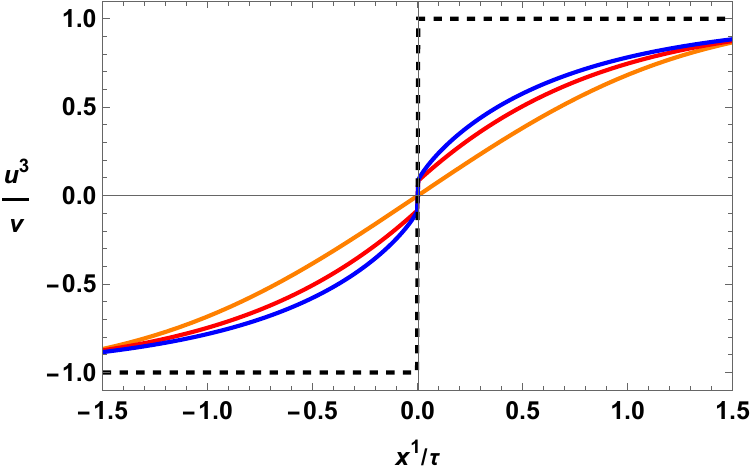}
\includegraphics[width=0.49\linewidth]{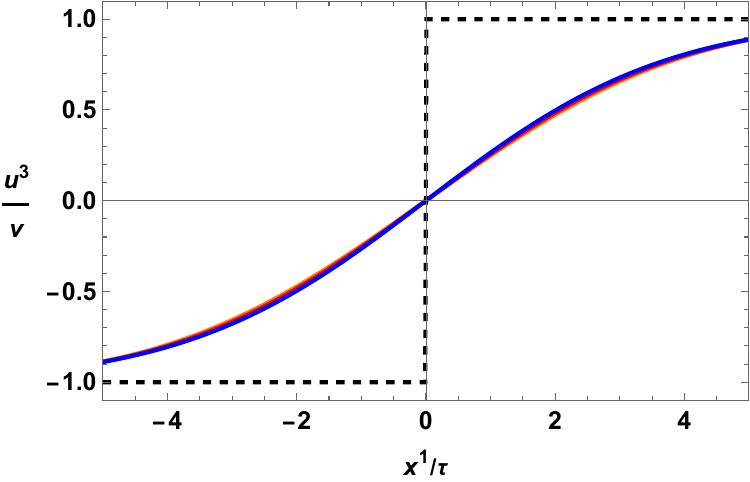}
    \caption{Radiative friction between two fluid layers of a grey material, with initial data (dashed black) as in figure \ref{fig:friction}. Each panel is a snapshot at a different time, respectively for $\eta^{(1)}t/W=0.05$ (up-left), $0.2$ (up-right), $0.5$ (down-left), and $5$ (down-right).  The exact solution of the linearised radiative Boltzmann equation (blue) is well reproduced by the non-Newtonian model \eqref{themodelone} (red) at all times, while it agrees with the Navier-Stokes solution (orange) only at late times.}
    \label{fig:exactevolution}
\end{figure}

An explicit example is provided in figure \ref{fig:exactevolution}, in the particular case of a grey material (for simplicity), i.e. a system such that $\tau_n(E){=}\text{const}{=}\tau{\in} \mathbb{R}$. In this limit, the flow, expressed in the coordinates $(\hat{x}^0\, ,\hat{x}^1){=}(\eta^{(1)}t/W \, ,x^1/\tau)$, is universal.

We stress that, while here the discussion was limited to the flow in figure \ref{fig:friction}, our results are more general. In fact, any linearized incompressible flow $\textbf{u}=\nabla \times \mathbb{A}$ (with $\mathbb{A}$ arbitrary) in 3D can be decomposed as a superposition of transversal plane waves, and each plane wave can be reoriented in space so that it fulfills \eqref{eom}, and thus decays according to \eqref{omegaalphato}. Therefore, our main conclusions hold for \textit{any} solenoidal velocity field in three dimensions. For illustration, in the Supplementary Material, we discuss the example of a vortex that decays due to radiative viscosity. The resulting trends are analogous to those in figure \ref{fig:exactevolution}: Navier-Stokes dramatically overestimates the early-time decay rate of optically thin structures, while our non-Newtonian model always maintains good accuracy.


\noindent\textit{\textbf{Mapping into Israel-Stewart -}} Along transversal flows, the linearised Israel-Stewart equations of motion read \cite{GavassinoNonHydro2022}
\begin{equation}
\begin{cases}
W \partial_t \delta u_3+ \partial_t \delta q_3 +\partial_1 \delta \Pi_{13}=0 \, ,\\
(\tau_q\partial_t {+}1)\delta q_3= -T\kappa_q (\partial_t \delta u_3-\alpha_1\partial_1 \delta\Pi_{13}) \, ,\\
(\tau_\Pi\partial_t {+}1)\delta \Pi_{13}= -\eta^{(1)} (\partial_1 \delta u_3-\alpha_1 \partial_1 \delta q_3) \, ,
\end{cases}
\end{equation}
for some transport coefficients $\tau_q$, $\tau_\Pi$ and $\alpha_1$. The dispersion relations are then roots of the polynomial (with $\lambda{\equiv}W^{-1}$)
\begin{equation}
(\lambda T\kappa_q-\tau_q )\tau_\Pi \omega^3 +i (\lambda T\kappa_q-\tau_\Pi -\tau_q) \omega^2 +\big(1+T\kappa_q \alpha_1^2\eta^{(1)}k^2+\lambda \eta^{(1)}k^2 (\tau_q{+}2T\kappa_q \alpha_1) \big)\omega+i \lambda\eta^{(1)} k^2 =0 \, .   
\end{equation}
In the limit of large $W$, the gapless solution can be found with the same rigorous strategy as in footnote \ref{bigproofs}, giving
\begin{equation}
\omega(k)=-\dfrac{i}{W} \, \dfrac{\eta^{(1)}k^2}{1+T\kappa_q \alpha_1^2 \eta^{(1)}k^2} \, ,
\end{equation}
which agrees with \eqref{limitsforti} provided that $\alpha_1^{-2}=\eta^{(-1)}T\kappa_q$. This completes our derivation of (i,ii,iii,iv).

\noindent\textit{\textbf{Discussion and Conclusions -}}  Navier-Stokes theory predicts that, if the gradients increase indefinitely, also the viscous stress $\Pi^{\text{NS}}_{13}$ increases indefinitely. Here, we proved that, in reality, the viscous stress begins to decrease past a characteristic gradient lengthscale, see equation \eqref{request}. Hence, just like we need flux limiters in models of radiative heat conduction, we also need ``viscosity limiters'' in models of radiative shear viscosity. However, such limiters are in Fourier space, and \textit{not} in real space. This means that, instead of magnitude cutoffs like $|\Pi_{13}|= \min \{|\Pi_{13}^{\text{NS}}|,\Pi_{\text{cutoff}}\}$, as in \cite{ShibataRadiation2011k}, we need more sophisticated ``smoothed out'' constitutive relations, like \eqref{nonNewt}, which suppress friction when the gradient lengthscale is shorter than the radiation mean free path. Israel-Stewart theory can reproduce this effect if we include the heat-viscosity coupling: As gradients increase, a heat flux $q_3 \propto \partial_1 \Pi_{13}$ is generated, which feedbacks on the stress through a term $\Pi_{13}^{\text{(heat)}} \propto \partial^2_1 \Pi_{13}$. The resulting constitutive relation is precisely equation \eqref{nonNewt}.


\noindent\textbf{\textit{Application 1: Photons -}} Let us calculate all infinite transport coefficients \eqref{etini} (including $\eta^{(-1)}$) explicitly for two specific systems. Our first example are photons, subject to absorption, emission, and isoenergetic isotropic scattering. In this case, equation \eqref{boltzmann} reads (see Supplementary Material)
\begin{equation}
\bigg(\partial_t {+}\dfrac{p^1}{p^0} \partial_1 \bigg) \delta f^\gamma=-(\sigma_A{+}\sigma_S) \bigg(\dfrac{df_{\text{BB}}}{dp^0} p^3 \delta u^3 +\delta f^\gamma \bigg) \, ,
\end{equation}
where $\{\sigma_A,\sigma_S,f_{BB} \}$ are the absorption coefficient, the scattering coefficient, and the black-body momentum distribution, all of which are functions of $p^0$ \cite[\S II.2]{Pomraning1973}. The resulting transport coefficients are
\begin{equation}
\eta^{(2a-1)}=\dfrac{4a_R T^4}{(1{+}2a)(3{+}2a)} \,  \dfrac{1}{\big\langle(\sigma_A{+}\sigma_S)^{2a-1} \big\rangle} \, ,
\end{equation}
where $T$ is the temperature, $a_R$ is the radiation constant, and $\langle ...\rangle$ is the Rosseland mean, defined e.g. in \cite[\S 6.7]{CastorRadiationBook_2004}.


\noindent\textbf{\textit{Application 2: Oscillating neutrinos -}} We consider a minimal quantum-kinetic model \cite{Vlasenko:2013fja} for neutrinos with two flavors, say $\{e,\mu\}$, that oscillate according to the toy Hamiltonian $H{=}\omega_0\sigma_x/4$ ($\omega_0{=}\text{const}$, and $\sigma_x$ is the Pauli matrix). If we account only for absorption and emission of $e$-neutrinos, with associated (grey) absorption coefficient $\tau^{-1}_e$, the linearized quantum Boltzmann equation reads \cite{Richers:2019grc}
\begin{equation}
\begin{split}
\bigg(\partial_t {+}\dfrac{p^1}{p^0} \partial_1 \bigg) \begin{bmatrix}
\delta f^e & \delta f^{x}{+}i\delta f^{y}  \\
\delta f^{x}{-}i\delta f^{y} & \delta f^\mu 
\end{bmatrix}
= -{}& \dfrac{1}{\tau_e} \dfrac{df_{\text{FD}}}{dp^0}  \begin{bmatrix}
p^3 \delta u^3 & 0 \\
0 & 0 \\
\end{bmatrix} \\
-{}& \dfrac{1}{2\tau_e}
\begin{bmatrix}
2\delta f^e & \delta f^{x}{+}i\delta f^{y}  \\
\delta f^{x}{-}i\delta f^{y} & 0
\end{bmatrix} \\
 -{}& \dfrac{\omega_0}{4}
 \begin{bmatrix}
2 \delta f^y & i(\delta f^\mu {-}\delta f^e) \\
i(\delta f^e{-}\delta f^\mu)& -2f^y
\end{bmatrix} \, ,\\
\end{split}    
\end{equation}
where $\delta f^x{+}i\delta f^y$ is the flavour quantum coherence, and $f_{\text{FD}}$ is the Fermi-Dirac distribution. Assuming for simplicity that the equilibrium state has vanishing neutrino chemical potential, we find
\begin{equation}
\eta^{(2a-1)}=\dfrac{4b_R T^4}{(1{+}2a)(3{+}2a)} \dfrac{(2\tau_e)^{2a-1}}{1{-}\varphi^2} \bigg[ \dfrac{1}{\big(1{+}\sqrt{1{-}\varphi^2}\big)^{2a-1}}+\dfrac{1}{\big(1{-}\sqrt{1{-}\varphi^2}\big)^{2a-1}}-2\varphi^2 \bigg]\, ,  
\end{equation}
where $b_R=7a_R/8$ is the Fermionic radiation constant, and $\varphi=\omega_0 \tau_e$ is the mean free oscillation phase.\\


\section*{Acknowledgements}

This work is supported by a Vanderbilt's Seeding Success Grant. I thank David Radice and Adam Burrows for reading the manuscript and providing useful comments.

\bibliography{Biblio}

\onecolumngrid
\newpage
\begin{center}
  \textbf{\large Non-Newtonian corrections to radiative viscosity: Israel-Stewart theory as a viscosity limiter\\ Supplementary Material}\\[.2cm]
  L. Gavassino\\[.1cm]
  {\itshape Department of Mathematics, Vanderbilt University, Nashville, TN, USA\\}
(Dated: \today)\\[1cm]
\end{center}

\setcounter{equation}{0}
\setcounter{figure}{0}
\setcounter{table}{0}
\setcounter{page}{1}
\renewcommand{\theequation}{S\arabic{equation}}
\renewcommand{\thefigure}{S\arabic{figure}}

\title{Gapless non-hydrodynamic modes in relativistic kinetic theory: Supplementary Material}
\author{L.~Gavassino}
\email{lorenzo.gavassino@vanderbilt.edu}
\affiliation{Department of Mathematics, Vanderbilt University, Nashville, TN 37211, USA}

\maketitle

\section{Universal behavior of the radiative transport equation for transversal flows}

The Boltzmann equation for an ideal massless gas mixture interacting with a medium can be decomposed as
\begin{equation}\label{sgrem}
p^\mu \partial_\mu f^A_{\textbf{p}} = \sum_{\text{``processes''}}  [\mathcal{C}^A_{\textbf{p}}]_\text{``process''}\, ,
\end{equation}
where each term $[\mathcal{C}^A_{\textbf{p}}]_\text{``process''}$ on the right-hand side is a collision integral associated with a certain interaction process. If we linearize this equation for a plane-wave perturbation $\delta \Psi(t,x^1)$ away from equilibrium, we obtain
\begin{equation}\label{whatIwant}
\bigg(\partial_t {+}\dfrac{p^1}{p^0} \partial_1 \bigg) \delta f^A_{\textbf{p}}= \sum_{\text{``processes''}}  \dfrac{1}{p^0} \, \delta [\mathcal{C}^A_{\textbf{p}}]_\text{``process''}
\end{equation}
In this section, we will show that, along transversal flows, the most common interaction processes are such that
\begin{equation}\label{griecz}
\dfrac{1}{p^0} \, \delta [\mathcal{C}^A_{\textbf{p}}]_\text{``process''} =[\mathcal{S}^A(p^0)]_{\text{``process''}} \,  p^3 \delta u^3 -\sum_B[\mathcal{M}\indices{^A _B}(p^0)]_{\text{``process''}} \, \delta f^B_{\textbf{p}} \, ,  
\end{equation}
for some related coefficients $[\mathcal{S}^A(p^0)]_{\text{``process''}}$ and $[\mathcal{M}\indices{^A _B}(p^0)]_{\text{``process''}}$. 

The relevant formulas for the interaction processes below are taken from \cite{Bruenn1985,Rampp2002,Levermore1984,ShibataRadiation2011k,Richers:2019grc}.

\subsection{Absorption and emission}

In the fluid's local rest frame, the collision integral due to absorption and emission of radiation by the fluid reads
\begin{equation}\label{absorp}
[\mathcal{C}^A_{\textbf{p}}]_\text{``abs''} =p^0\bigg(\kappa_A(p^0)-\sigma_{A,a}(p^0)f^A_{\textbf{p}} \bigg) \, ,
\end{equation}
where $\kappa_A$ and $\sigma_{A,a}$ are the emission and absorption coefficients (already corrected with final-state terms). Since the collision integral and the distribution function are Lorentz scalars, equation \eqref{absorp} corresponds to the following covariant expression, which is true in all reference frames:
\begin{equation}\label{absorp2}
[\mathcal{C}^A_{\textbf{p}}]_\text{``abs''} =-u_\mu p^\mu\bigg(\kappa_A(-u_\mu p^\mu)-\sigma_{A,a}(-u_\mu p^\mu)f^A_{\textbf{p}} \bigg) \, .
\end{equation}
Linearizing this expression for a transversal flow (as discussed in the main text), we obtain
\begin{equation}\label{sgrom}
\begin{split}
\delta [\mathcal{C}^A_{\textbf{p}}]_\text{``abs''} ={}& -p^3 \delta u^3 \bigg(\kappa_A(p^0)-\sigma_{A,a}(p^0)f^A_{\textbf{p}} \bigg) \\
& - p^0 p^3 \delta u^3 \bigg(\dfrac{d\kappa_A(p^0)}{dp^0}-\dfrac{d \sigma_{A,a}(p^0)}{dp^0}f^A_{\textbf{p}} \bigg)\\
& -p^0 \sigma_{A,a}(p^0) \delta f^A_{\textbf{p}} \, , \\
\end{split}
\end{equation}
where all the quantities without ``$\delta$'' are evaluated in equilibrium. Now, let us note that, in global equilibrium, the Kirchhoff-Planck relation holds, namely $\kappa_A=\sigma_{A,a}f^A_{\textbf{p}}$, and $f^A_{\textbf{p}}$ depends only on $p^0$. Hence, \eqref{sgrom} simplifies to
\begin{equation}\label{asumptzio}
 \dfrac{1}{p^0} \delta [\mathcal{C}^A_{\textbf{p}}]_\text{``abs''} = -\sigma_{A,a}(p^0) \bigg(\dfrac{df^A_{p^0}}{dp^0} p^3 \delta u^3 +\delta f^A_\textbf{p}\bigg)\, , 
\end{equation}
which has indeed the form \eqref{griecz}, with $\mathcal{M}\indices{^A _B}\propto \delta\indices{^A _B}$.

\subsection{Isoenergetic scattering}

If a scattering process (of a radiation particle against matter) conserves the energy of the scattered radiation, and is isotropic, the relative contribution to the Boltzmann collision integral takes the following form:
\begin{equation}
\begin{split}
[\mathcal{C}^A_{\textbf{p}}]_{\text{``scat''}} ={}& p^0  \sigma_{A,s}(p^0)\int_{\mathcal{S}^2} \dfrac{d^2\Omega'}{4\pi} (f^A_{p^0\mathbf{\Omega}'}-f^A_{\textbf{p}}) \\ 
={}& \sigma_{A,s}(p^0) \int_{\mathcal{S}^2} \dfrac{d^2\Omega'}{4\pi} \int_0^{\infty} p'^0 dp'^0 \delta(p'^0{-} p^0) \,  (f^A_{p'^0\mathbf{\Omega}'}-f^A_{\textbf{p}}) \, ,
\end{split}
\end{equation}
where $\sigma_{A,s}$ is the scattering coefficient, $d^2 \Omega$ is the solid angle element, and $\mathbf{\Omega}$ is a unit vector.
Recalling that the volume element in spherical coordinates is $d^3p=(p'^0)^2 dp'^0 d^2\Omega$, we can rewrite the above integral as follows:
\begin{equation}\label{in3d}
[\mathcal{C}^A_{\textbf{p}}]_{\text{``scat''}} = \sigma_{A,s}(p^0)\int_{\mathbb{R}^3} \dfrac{d^3 p'}{4\pi p'^0} \delta(p'^0{-}p^0)  (f^A_{\textbf{p}'} -f^A_{\textbf{p}}  ) 
\end{equation}
Hence, recalling that $d^3p/p^0$ is the invariant measure in momentum space, equation \eqref{in3d} corresponds to the following Lorentz-covariant expression, valid in all inertial frames:
\begin{equation}
\begin{split}
[\mathcal{C}^A_{\textbf{p}}]_{\text{``scat''}} ={}& \sigma_{A,s}(-u_\mu p^\mu) \int_{\mathbb{R}^3} \dfrac{d^3 p'}{4\pi p'^0} \delta(-u_\mu p'^\mu{+}u_\mu p^\mu)  (f^A_{\textbf{p}'}-f^A_{\textbf{p}}) \\
={}& \sigma_{A,s}(-u_\mu \Omega^\mu p^0) \int_{\mathbb{R}^3} \dfrac{d^3 p'}{4\pi p'^0} \delta(-u_\mu \Omega'^\mu p'^0{+}u_\mu \Omega^\mu p^0)  (f^A_{\textbf{p}'}-f^A_{\textbf{p}}) \\
={}& \sigma_{A,s}(-u_\mu \Omega^\mu p^0)\int_{\mathcal{S}^2} \dfrac{d^2\Omega'}{4\pi} \int_0^{\infty} \dfrac{p'^0 dp'^0}{-u_\mu \Omega'^\mu} \delta \bigg( p'^0{-}\dfrac{u_\mu \Omega^\mu}{u_\mu \Omega'^\mu} p^0 \bigg)  (f^A_{p'^0\mathbf{\Omega}'}-f^A_{\textbf{p}}) \\
={}& p^0 \sigma_{A,s}(-u_\mu \Omega^\mu p^0) \int_{\mathcal{S}^2} \dfrac{d^2\Omega'}{4\pi}  \dfrac{ -u_\mu \Omega^\mu}{(-u_\mu \Omega'^\mu)^2}  \bigg(f^A_{\frac{u_\mu \Omega^\mu}{u_\mu \Omega'^\mu} p^0\mathbf{\Omega}'}-f^A_{\textbf{p}} \bigg) \\
\end{split}
\end{equation}
where we introduced the notation $\Omega^\mu=(1,\mathbf{\Omega})=p^\mu/p^0$. In our case, the motion is confined in the $x^3$ direction, and the velocity is very small, so that $u^\mu \approx (1,0,0,u^3)$. Hence, we can write
\begin{equation}
\dfrac{1}{p^0}[\mathcal{C}^A_{\textbf{p}}]_{\text{``scat''}} =\sigma_{A,s}\big((1-u_3 \Omega^3) p^0 \big) \int_{\mathcal{S}^2} \dfrac{d^2 \Omega'}{4\pi}  \dfrac{1{-}u_3\Omega^3}{(1{-}u_3 \Omega'^3)^2} \, \bigg(   f^A_{\frac{1{-}u_3 \Omega^3}{1-u_3 \Omega'^3} p^0\mathbf{\Omega}'} -f^A_{\textbf{p}}\bigg)
\end{equation}
Now we only need to expand this formula to linear order in deviations of $u^3$ and $f^A_\textbf{p}$ from their equilibrium values. The result is provided below:
\begin{equation}
\begin{split}
\dfrac{1}{p^0} \delta [\mathcal{C}^A_{\textbf{p}}]_{\text{``scat''}} ={}& -p^3 \delta u^3 \dfrac{d \sigma_{A,s}(p^0)}{dp^0} \int_{\mathcal{S}^2} \dfrac{d^2 \Omega'}{4\pi}  \,   (f^A_{ p^0\mathbf{\Omega}'}-f^A_{\textbf{p}}) \\
+{}&  \sigma_{A,s}(p^0) \int_{\mathcal{S}^2} \dfrac{d^2 \Omega'}{4\pi}  (2\Omega'^3-\Omega^3)\delta u^3 \,   (f^A_{ p^0\mathbf{\Omega}'}-f^A_{\textbf{p}}) \\
+{}& \sigma_{A,s}(p^0) \int_{\mathcal{S}^2} \dfrac{d^2 \Omega'}{4\pi}   \,  \dfrac{d f^A_{p^0}}{d p^0} (\Omega'^3-\Omega^3)p^0 \delta u^3 \\
+{}& \sigma_{A,s}(p^0) \int_{\mathcal{S}^2} \dfrac{d^2 \Omega'}{4\pi}   \,  (\delta f^A_{p^0\mathbf{\Omega}'}-\delta f^A_{\textbf{p}}) \, ,\\
\end{split}
\end{equation}
where $f^A_{\textbf{p}}=f^A_{p^0\mathbf{\Omega}} \equiv f^A_{p^0}$ is the equilibrium distribution, which depends only on the energy. Clearly, the first and second lines vanish (because $f^A_{ p^0\mathbf{\Omega}'}-f^A_{\textbf{p}}$ is identically zero), and the term $\Omega'^3$ in the third line averages to zero. In the fourth line, the integral $\int d^2 \Omega \, \delta f^A_{p^0\mathbf{\Omega}'}$ vanishes identically, because the function $\delta f^A_{p^0\mathbf{\Omega}'}$ is odd under the 180$^o$-degree rotation $(\Omega'^1,\Omega'^2,\Omega'^3)\rightarrow (\Omega'^1,-\Omega'^2,-\Omega'^3)$, as discussed in the main text. Hence, we are left with
\begin{equation}
\dfrac{1}{p^0} \delta [\mathcal{C}^A_{\textbf{p}}]_{\text{``scat''}}=- 
 \sigma_{A,s}(p^0)   \bigg[  \dfrac{d f^A_{p^0}}{d p^0} p^3 \delta u^3 +  \delta f^A_{\textbf{p}} \bigg] \, ,
\end{equation}
which is formally identical to \eqref{asumptzio}.

\subsection{Scattering: more in general}\label{scattone}

Let us now consider a more general type of scattering process, which does not conserve the rest-frame energy, but is still isotropic. Then,
in the local rest frame of the medium, the collision integral reads
\begin{equation}\label{scatt1}
\dfrac{1}{p^0}[\mathcal{C}^A_{\textbf{p}}]_{\text{``scat''}}= (1{+}z_A f^A_{\textbf{p}})\int \dfrac{d^3 p'}{(2\pi)^3} f^A_{\textbf{p}'} \,  R^{\text{in}}_A(p^0,p'^0)-   f^A_{\textbf{p}}\int \dfrac{d^3 p'}{(2\pi)^3} (1{+}z_Af^A_{\textbf{p}'} )\, R^{\text{out}}_A(p^0,p'^0) \, ,
\end{equation}
where $R^{\text{in}}_A$ and $R^{\text{out}}_A$ are the scattering kernels, and $z_A$ is the statistics identifier for the radiaiton species $A$, which equals $+1$ for Bosons, $-1$ for Fermions, and $0$ for for non-degenerate particles. Note that, apart from the isotropy assumption, equation \eqref{scatt1} is completely general, and is valid both for photons and neutrinos.

For our purposes, it is more convenient to rewrite \eqref{scatt1} as follows:
\begin{equation}\label{scatt2}
[\mathcal{C}^A_{\textbf{p}}]_{\text{``scat''}}= (1{+}z_A f^A_{\textbf{p}})\int \dfrac{d^3 p'}{(2\pi)^3p'^0} f^A_{\textbf{p}'} \, \tilde{R}^{\text{in}}_A(p^0,p'^0)-   f^A_{\textbf{p}}\int \dfrac{d^3 p'}{(2\pi)^3p'^0} (1{+}z_Af^A_{\textbf{p}'} )\, \tilde{R}^{\text{out}}_A(p^0,p'^0) \, ,
\end{equation}
with $\tilde{R}^{\text{in}}_A=p^0 p'^0 R^{\text{in}}_A$ and $\tilde{R}^{\text{out}}_A=p^0 p'^0 R^{\text{out}}_A$. Then, recalling that the measure $d^3p'/p'^0$ is Lorentz invariant, equation \eqref{scatt2} correspond to the following covariant expression, which is valid in all reference frames:
\begin{equation}\label{scatt3}
[\mathcal{C}^A_{\textbf{p}}]_{\text{``scat''}}= (1{+}z_A f^A_{\textbf{p}})\int \dfrac{d^3 p'}{(2\pi)^3p'^0} f^A_{\textbf{p}'} \, \tilde{R}^{\text{in}}_A(-u_\mu p^\mu,-u_\mu p'^\mu)-   f^A_{\textbf{p}}\int \dfrac{d^3 p'}{(2\pi)^3p'^0} (1{+}z_Af^A_{\textbf{p}'} )\, \tilde{R}^{\text{out}}_A(-u_\mu p^\mu,-u_\mu p'^\mu) \, .
\end{equation}
Linearizing this expression along a transversal flow, as we did before, we obtain the following:
\begin{equation}
\begin{split}
\delta [\mathcal{C}^A_{\textbf{p}}]_{\text{``scat''}} &{}= \delta f^A_{\textbf{p}} \bigg[ z_A\int \dfrac{d^3 p'}{(2\pi)^3p'^0} f^A_{\textbf{p}'} \, \tilde{R}^{\text{in}}_A(p^0,p'^0)-  \int \dfrac{d^3 p'}{(2\pi)^3p'^0} (1{+}z_Af^A_{\textbf{p}'} )\, \tilde{R}^{\text{out}}_A(p^0,p'^0)\bigg] \\ 
& +(1{+}z_A f^A_{\textbf{p}})\int \dfrac{d^3 p'}{(2\pi)^3p'^0} \delta f^A_{\textbf{p}'} \, \tilde{R}^{\text{in}}_A(p^0,p'^0)- z_A   f^A_{\textbf{p}}\int \dfrac{d^3 p'}{(2\pi)^3p'^0} \delta f^A_{\textbf{p}'} \, \tilde{R}^{\text{out}}_A(p^0,p'^0) \\ 
&-p^3 \delta u^3 \bigg[(1{+}z_A f^A_{\textbf{p}})\int \dfrac{d^3 p'}{(2\pi)^3p'^0} f^A_{\textbf{p}'} \, \partial_{p^0}\tilde{R}^{\text{in}}_A(p^0,p'^0)-   f^A_{\textbf{p}}\int \dfrac{d^3 p'}{(2\pi)^3p'^0} (1{+}z_Af^A_{\textbf{p}'} )\, \partial_{p^0}\tilde{R}^{\text{out}}_A(p^0,p'^0)\bigg]\\
&- \delta u^3 \bigg[(1{+}z_A f^A_{\textbf{p}})\int \dfrac{d^3 p'}{(2\pi)^3p'^0} f^A_{\textbf{p}'}\, p'^3 \, \partial_{p'^0}\tilde{R}^{\text{in}}_A(p^0,p'^0)-   f^A_{\textbf{p}}\int \dfrac{d^3 p'}{(2\pi)^3p'^0} (1{+}z_Af^A_{\textbf{p}'} )\, p'^3 \, \partial_{p'^0}\tilde{R}^{\text{out}}_A(p^0,p'^0)\bigg]\, ,\\
\end{split}
\end{equation}
where, again, all the terms without ``$\delta$'' are evaluated in equilibrium. Now, we note that, in the fourth line, the integrals average to zero, because the functions $f^A$ and $\Tilde{R}_A$ are isotropic, while $p'^3$ is odd. Indeed, in a transversal flow, also the second line vanishes, because the perturbation $\delta f^A_\textbf{p}$ is odd under transformation $(p^1,p^2,p^3)\rightarrow (p^1,-p^2,-p^3)$. Therefore, we have
\begin{equation}
\begin{split}
\dfrac{1}{p^0} \delta[\mathcal{C}^A_{\textbf{p}}]_{\text{``scat''}} &{}= \delta f^A_{\textbf{p}} \bigg[ z_A\int \dfrac{d^3 p'}{(2\pi)^3} f^A_{\textbf{p}'} \, R^{\text{in}}_A(p^0,p'^0)-  \int \dfrac{d^3 p'}{(2\pi)^3} (1{+}z_Af^A_{\textbf{p}'} )\, R^{\text{out}}_A(p^0,p'^0)\bigg] \\ 
&-p^3 \delta u^3 \bigg[(1{+}z_A f^A_{\textbf{p}})\int \dfrac{d^3 p'}{(2\pi)^3p^0 p'^0} f^A_{\textbf{p}'} \, \partial_{p^0}\tilde{R}^{\text{in}}_A(p^0,p'^0)-   f^A_{\textbf{p}}\int \dfrac{d^3 p'}{(2\pi)^3 p^0 p'^0} (1{+}z_Af^A_{\textbf{p}'} )\, \partial_{p^0}\tilde{R}^{\text{out}}_A(p^0,p'^0)\bigg] \, ,\\
\end{split}
\end{equation}
which has again the form \eqref{griecz}, with $\mathcal{M}\indices{^A _B}\propto \delta\indices{^A _B}$.

\subsection{Pair production and annihilation}

Neutrinos can be emitted or absorbed in couples, throw pair processes: $l^+{+}l^- \ce{<=>} \nu {+}\Bar{\nu}$ (for some lepton $l$). Assuming isotropy, the related collision integral in the fluid's rest frame reads ($A=$ neutrino, $B=$ antineutrino)
\begin{equation}\label{pair1}
\dfrac{1}{p^0}[\mathcal{C}^A_{\textbf{p}}]_{\text{``pair''}}= \int \dfrac{d^3 p'}{(2\pi)^3} \bigg[(1{+}z_A f^A_\textbf{p})(1{+}z_B f^B_{\textbf{p}'})R^{\text{pro}}_{AB}(p^0,p'^0) -f^A_\textbf{p}f^B_{\textbf{p}'} R^{\text{ann}}_{AB}(p^0,p'^0) \bigg] \, ,
\end{equation}
for some collision kernels $R^{\text{pro}}_{AB}$ and $R^{\text{ann}}_{AB}$. Like with scattering, we can rewrite the integral in a more convenient form:
\begin{equation}\label{pair2}
[\mathcal{C}^A_{\textbf{p}}]_{\text{``pair''}}= \int \dfrac{d^3 p'}{(2\pi)^3p'^0} \bigg[(1{+}z_A f^A_\textbf{p})(1{+}z_B f^B_{\textbf{p}'})\Tilde{R}^{\text{pro}}_{AB}(p^0,p'^0) -f^A_\textbf{p}f^B_{\textbf{p}'} \Tilde{R}^{\text{ann}}_{AB}(p^0,p'^0) \bigg] \, ,
\end{equation}
with $\tilde{R}=p^0p'^0 R$. Boosting to a generic reference frame, we obtain
\begin{equation}\label{pair3}
[\mathcal{C}^A_{\textbf{p}}]_{\text{``pair''}}= \int \dfrac{d^3 p'}{(2\pi)^3p'^0} \bigg[(1{+}z_A f^A_\textbf{p})(1{+}z_B f^B_{\textbf{p}'})\Tilde{R}^{\text{pro}}_{AB}(-u_\mu p^\mu,-u_\mu p'^\mu) -f^A_\textbf{p}f^B_{\textbf{p}'} \Tilde{R}^{\text{ann}}_{AB}(-u_\mu p^\mu,-u_\mu p'^\mu) \bigg] \, .
\end{equation}
Linearizing on a transversal flow, we get
\begin{equation}
\begin{split}
\delta [\mathcal{C}^A_{\textbf{p}}]_{\text{``pair''}} &{}= \delta f^A_\textbf{p} \int \dfrac{d^3 p'}{(2\pi)^3p'^0} \bigg[z_A (1{+}z_B f^B_{\textbf{p}'})\Tilde{R}^{\text{pro}}_{AB}(p^0,p'^0) -f^B_{\textbf{p}'} \Tilde{R}^{\text{ann}}_{AB}(p^0,p'^0) \bigg] \\
&{}+ \int \dfrac{d^3 p'}{(2\pi)^3p'^0} \bigg[(1{+}z_A f^A_\textbf{p})z_B \delta f^B_{\textbf{p}'}\Tilde{R}^{\text{pro}}_{AB}(p^0,p'^0) -f^A_\textbf{p} \delta f^B_{\textbf{p}'} \Tilde{R}^{\text{ann}}_{AB}(p^0,p'^0) \bigg] \\
&{}-p^3 \delta u^3  \int \dfrac{d^3 p'}{(2\pi)^3p'^0} \bigg[(1{+}z_A f^A_\textbf{p})(1{+}z_B f^B_{\textbf{p}'})\partial_{p^0}\Tilde{R}^{\text{pro}}_{AB}(p^0,p'^0) -f^A_\textbf{p}f^B_{\textbf{p}'} \partial_{p^0}\Tilde{R}^{\text{ann}}_{AB}(p^0,p'^0) \bigg] \\
&{}- \delta u^3  \int \dfrac{d^3 p'}{(2\pi)^3p'^0} \bigg[(1{+}z_A f^A_\textbf{p})(1{+}z_B f^B_{\textbf{p}'})p'^3\partial_{p'^0}\Tilde{R}^{\text{pro}}_{AB}(p^0,p'^0) -f^A_\textbf{p}f^B_{\textbf{p}'} p'^3\partial_{p'^0}\Tilde{R}^{\text{ann}}_{AB}(p^0,p'^0) \bigg] \, .\\
\end{split}
\end{equation}
As in the scattering case, the second line vanishes on transversal flows, because the perturbation $\delta f^B_{\textbf{p}'}$ is odd under the transformation $(p^1,p^2,p^3)\rightarrow (p^1,-p^2,-p^3)$. The fourth line also vanishes because $p'^3$ is an odd function, while $f^B_{\textbf{p}}$ is isotropic (being the equilibrium distribution). Thus, we finally have
\begin{equation}
\begin{split}
\delta \dfrac{1}{p^0}[\mathcal{C}^A_{\textbf{p}}]_{\text{``pair''}} &{}= \delta f^A_\textbf{p} \int \dfrac{d^3 p'}{(2\pi)^3} \bigg[z_A (1{+}z_B f^B_{\textbf{p}'})R^{\text{pro}}_{AB}(p^0,p'^0) -f^B_{\textbf{p}'} R^{\text{ann}}_{AB}(p^0,p'^0) \bigg] \\
&{}-p^3 \delta u^3  \int \dfrac{d^3 p'}{(2\pi)^3 p^0p'^0} \bigg[(1{+}z_A f^A_\textbf{p})(1{+}z_B f^B_{\textbf{p}'})\partial_{p^0}\Tilde{R}^{\text{pro}}_{AB}(p^0,p'^0) -f^A_\textbf{p}f^B_{\textbf{p}'} \partial_{p^0}\Tilde{R}^{\text{ann}}_{AB}(p^0,p'^0) \bigg] \, , \\
\end{split}
\end{equation}
which again has the form \eqref{griecz}, with $\mathcal{M}\indices{^A _B}\propto \delta\indices{^A _B}$.

\subsection{Flavour oscillation}

As they travel across spacetime, neutrinos can spontaneously change flavor, in a process called ``neutrino oscillation''. Strictly speaking, this peculiar type of chemical transformation is not the result of a collision process. However, we may still include its effect on the right-hand side of \eqref{whatIwant}, as an additional process. Let us see how this work.

To account for neutrino oscillations, we must promote $f^A_{\textbf{p}}$ to a quantum coherence matrix. This means that $A$ is now a double index, $A=(a,b)$, where $a$ and $b$ run over all the allowed flavor states. For standard-model neutrinos, one has $a,b\in\{e,\mu,\tau \}$, and the distribution function may be written as follows:
\begin{equation}
f^A_{\textbf{p}}=f^{ab}_{\textbf{p}}=
\begin{bmatrix}
f^{ee}_{\textbf{p}} & f^{e\mu}_{\textbf{p}} & f^{e\tau}_{\textbf{p}} \\
f^{\mu e}_{\textbf{p}} & f^{\mu \mu}_{\textbf{p}} & f^{\mu \tau}_{\textbf{p}} \\
f^{\tau e}_{\textbf{p}} & f^{\tau \mu}_{\textbf{p}} & f^{\tau \tau}_{\textbf{p}} \\
\end{bmatrix} \, .
\end{equation}
For neutrinos freely traveling in vacuum, one has, in a generic reference frame,
\begin{equation}
\bigg(\partial_t + \dfrac{p^j}{p^0} \partial_j \bigg) f^A_{\textbf{p}} =-i[H(p^0),f_{\textbf{p}}]^A \, ,
\end{equation}
where $H(p^0)$ is the Hamiltonian governing the oscillations. This has indeed the form \eqref{whatIwant}, with
\begin{equation}
    \dfrac{1}{p^0} \, \delta [\mathcal{C}^A_{\textbf{p}}]_\text{``osc''} =-i[H(p^0),\delta f_{\textbf{p}}]^A\, .
\end{equation}
Since the entries of $[H,\delta f]$ are linear combinations of the entries of $\delta f$, there exists a matrix $[\mathcal{M}\indices{^A _B}(p^0)]_{\text{``osc''}}$ such that
\begin{equation}
    \dfrac{1}{p^0} \, \delta [\mathcal{C}^A_{\textbf{p}}]_\text{``osc''} =-\sum_B [\mathcal{M}\indices{^A _B}(p^0)]_{\text{``osc''}} \, \delta f_{\textbf{p}}^B\, ,  
\end{equation}
which has indeed the form \eqref{griecz}

\newpage

\section{Shear waves with M1 closure cannot decay}

In this section, we prove that current simulations with M1 closure do not include an actual radiative shear viscosity.

To avoid possible misunderstandings, let us state precisely what we mean by that. To have ``shear viscosity'', a hydrodynamic theory does not need to have an explicit gradient correction to the stress tensor, proportional to $\partial_{1}u_{3}$. Any form of shear momentum flux $T_{13}$ may play the role of an effective shear viscosity, if it causes friction between two layers in relative motion. However, for this flux to act as a genuine shear viscosity, we need a mechanism that will ``turn on'' this shear stress in the presence of \textit{any} shear flow. In other words, we need some term in the equation of motion (of whatever form) that ``activates'' $T_{13}$ when the flow has a geometry as in figure 1 of the main text. Here, we show that the M1 closure does not have such a term. In particular, we prove that there exist exact stationary solutions of M1 radiation hydrodynamics where the fluid layers are (and forever will be) experiencing shear motion, and they keep sliding against each other with no friction.

\vspace{-0.3cm}
\subsection{Overview of the M1 closure scheme}
\vspace{-0.2cm}

In radiation-hydrodynamic simulations, the M1 closure scheme \cite{Minerbo1978,Levermore1984,Sadowski2013,Fragile:2014bfa,Fragile:2018xee,Murchikova:2017zsy,Anninos2020,RadiceNew2022,GavassinoRadiazione} is an approximate method that reduces the number of dynamical degrees of freedom of the radiation component. It does not track the evolution of the whole distribution function $f^A_\textbf{p}(x^\alpha)$ explicitly. Instead, it only follows the evolution of two effective fields, $\{\varepsilon_R(x^\alpha),F^\nu(x^\alpha)\}$, which represent respectively the radiation energy density and the radiation energy flux in the rest frame of the medium (note that $F^\nu u_\nu =0$). The radiation stress-energy tensor is then assumed to be 
\begin{equation}\label{dedkorfc}
T_\text{rad}^{\mu \nu}= \dfrac{4}{3} \varepsilon_R u^\mu u^\nu +\dfrac{1}{3}\varepsilon_R g^{\mu \nu } +F^\mu u^\nu + u^\mu F^\nu+ \dfrac{3\chi{-}1}{2} \varepsilon_R \bigg[ \dfrac{F^\mu F^\nu}{F^\alpha F_\alpha} {-}\dfrac{g^{\mu \nu} {+}u^\mu u^\nu}{3} \bigg] \, , 
\end{equation}
where the last term models anisotropies in the pressure due to the presence of a radiative flux $F^\nu$. The assumption that such an effect can be parameterized in the above form is the central approximation of the scheme. The quantity $\chi$ is the so-called ``Eddington factor'' \cite{Minerbo1978,Levermore1984}, which is taken to be a function of the scalar $F^\alpha F_\alpha/\varepsilon_R^2$. For small $F^\nu$, we can write down an expansion of the form $\chi \approx 1/3 +z \, F^\alpha F_\alpha/\varepsilon_R^2$, for some number $z$. Hence, in the linear regime, the pressure anisotropy term vanishes.

The equation of motion for $\varepsilon_R$ and $F^\nu$ is a balance law of the form:
\begin{equation}\label{basilisco}
\partial_\mu T_\text{rad}^{\mu \nu}= - (\varepsilon_R -\varepsilon_{R}^\text{LTE}) G_1 u^\nu -G_2 F^\nu \, ,
\end{equation}
for some coefficients $G_1$ and $G_2$, which incorporate all the relevant interaction processes. The quantity $\varepsilon_{R}^\text{LTE}=\varepsilon_{R}^\text{LTE}(\Lambda)$ is just the value of $\varepsilon_{R}$ in local thermodynamic equilibrium. 

\vspace{-0.3cm}
\subsection{Proof in the linear limit}
\vspace{-0.2cm}

We recall that, along linear transversal flows, the function $\delta f_\text{p}^A$ is odd under the rotation $(p^1,p^2,p^3)\rightarrow (p^1,-p^2,-p^3)$. Thus, we have that
\begin{equation}
\delta \varepsilon_R = \sum_A g_A \int \dfrac{d^3 p}{(2\pi)^3} p^0 \delta f^A_\textbf{p}=0 \, , \spc \spc
\delta F^1 = \sum_A g_A \int \dfrac{d^3 p}{(2\pi)^3} p^1 \delta f^A_\textbf{p}=0 \, . 
\end{equation}
Assuming, as is done in Figure 1 of the main text, that the flow points in direction $x^3$, we can also set $\delta F^2=0$. Hence, the only relevant degrees of freedom of the transversal flow are $\{\delta u^3,\delta F^3 \}$. Therefore, the M1 analog of equation (7) of the main text is 
\begin{subequations}
		\label{eomM1}
		\begin{eqnarray}
&\label{conservationM1} \bigg(W+\dfrac{4}{3}\varepsilon_R \bigg) \partial_t \delta u^3 + \partial_t \delta F^3=0 \, , \\
&\label{boltzmannM1} \dfrac{4}{3}\varepsilon_R \partial_t \delta u^3 +\partial_t \delta F^3 = -G_2 \delta F^3 \, .
\end{eqnarray}
\end{subequations}
Again, the first equation is just the conservation of linear momentum, $\partial_\mu \delta T^{\mu 3}_{\text{fluid}}{+}\partial_\mu \delta T^{\mu 3}_{\text{rad}}{=}0$, while the second equation is the linearization of \eqref{basilisco}. Now we immediately see the problem: All configurations of the form
\begin{equation}
\begin{cases}
\delta u^3(t,x^1)=v(x^1) \, , \\
\delta F^3(t,x^1)=0  \\
\end{cases}
\end{equation}
are solutions of \eqref{eomM1}, for any choice of function $v$. This tells us that shear waves survive \textit{forever}, and the (effective) shear viscosity coefficient is identically zero. Indeed, if we take $v(x^1)=v_0 \, \text{sgn} (x^1)$, as in figure 1 of the main text, we find that there is no friction between the fluid layers, since $\delta u^3$ does not decrease over time. 

\subsection{Proof in the non-linear regime}

Let us now consider the full non-linear problem. The exact M1 equations of motion are the following:
\begin{subequations}
		\label{eomM13}
		\begin{eqnarray}
&\label{conservationM133} \partial_\mu (n_I u^{\mu})=R_I \, , \\
&\label{conservationM13} \partial_\mu T_\text{fluid}^{\mu \nu}= (\varepsilon_R -\varepsilon_{R}^\text{LTE}) G_1 u^\nu +G_2 F^\nu \, , \\
&\label{boltzmannM13} \partial_\mu T_\text{rad}^{\mu \nu}= - (\varepsilon_R -\varepsilon_{R}^\text{LTE}) G_1 u^\nu -G_2 F^\nu \, .
\end{eqnarray}
\end{subequations}
Equation \eqref{conservationM133} is the continuity equation for all chemical constituents $I$ of matter, with some reaction rates $R_I$. Equation \eqref{boltzmannM13} is the same as equation \eqref{basilisco}, while equation \eqref{conservationM13} follows from the conservation laws of energy and momentum, namely $\partial_\mu (T_\text{fluid}^{\mu \nu}{+}T_\text{rad}^{\mu \nu})=0$. Let us now prove that the stationary ``eternal'' shear wave
\begin{equation}\label{statesloro}
\begin{cases}
u^\mu = \dfrac{1}{\sqrt{1{-}v(x^1)^2}}\big(1,0,0,v(x^1)\big)\, , \\
F^\mu =0 \, , \\
(\Lambda,\varepsilon_R) =\text{const}=\text{``local thermodynamic equilibrium''}  \\
\end{cases}
\end{equation}
is an \textit{exact} analytical solution of the system \eqref{eomM13}, for any choice of profile $v(x^1)\in (-1,1)$. To this end, we note that, since matter and radiation are in local thermodynamic equilibrium (which includes chemical equilibrium), the right-hand sides of \eqref{eomM13} vanish identically. Furthermore, since $\varepsilon_R =\varepsilon_{R}^\text{LTE}(\Lambda)$ and $F^\mu{=}0$, equation \eqref{dedkorfc} simplifies to
\begin{equation}
T_\text{rad}^{\mu \nu}= \dfrac{4}{3} \varepsilon_{R}^\text{LTE} u^\mu u^\nu +\dfrac{1}{3} \varepsilon_{R}^\text{LTE} g^{\mu \nu }  \, .
\end{equation}
Therefore, recalling that all fields depend only $x^1$, the system \eqref{eomM13} becomes
\begin{subequations}
		\label{eomM134}
		\begin{eqnarray}
&\label{conservationM1334} \partial_1 (n_I u^{1})=0 \, , \\
&\label{conservationM134} \partial_1 \big[ W u^1 u^\nu +P_\text{fluid} g^{1\nu} \big]=0 \, , \\
&\label{boltzmannM134} \partial_1 \bigg[ \dfrac{4}{3} \varepsilon_{R}^\text{LTE} u^1 u^\nu +\dfrac{1}{3} \varepsilon_{R}^\text{LTE} g^{1\nu } \bigg]=0 \, .
\end{eqnarray}
\end{subequations}
Now we only need to notice that, by assumption, $u^1=0$ and $g^{\mu \nu}=\text{diag}(-1,1,1,1)$. With this geometry, the system \eqref{eomM134} is satisfied, provided that $\partial_1 P_\text{fluid}=\partial_1 \varepsilon_{R}^\text{LTE}=0$, which is indeed true in the configurations \eqref{statesloro}. This completes our proof. 

We stress that, while the states \eqref{statesloro} solve the M1 equations of motion, they \textit{do not} solve the full Boltzmann equation unless $v=\text{const}$. In fact, the assumption of local thermodynamic equilibrium implies that (a) $f^A_\textbf{p}=f^A_\text{LTE}(-u_\nu p^\nu)$, and (b) the collision integral must vanish. Hence, the transport equation becomes $p^\mu \partial_\mu f^A_\text{LTE}(-u_\nu p^\nu)=0$, which gives
\begin{equation}
0= p^\mu \partial_\mu f^A_\text{LTE}(-u_\nu p^\nu) = p^1 \partial_1 f^A_\text{LTE}( u^0 p^0{-}u^3 p^3)={f^A_\text{LTE}}' \,  p^1 \partial_1( u^0 p^0{-}u^3 p^3) \, .
\end{equation}
This condition is obeyed for all choices of $(p^1,p^2,p^3)$ only if $v(x^1)=\text{const}$, i.e. in the absence of shear waves. This shows that, while M1 closure does not have a shear viscosity, the full Boltzmann equation has it.

\subsection{Multi-frequency radiation hydrodynamics with M1 closure}

For completeness, let us also discuss the ``multi-frequency'' method. The latter is a refinement of M1 radiation hydrodynamics, where one does not just track the total radiation energy density and flux. Instead, one separately evolves a large number of radiation energy densities and fluxes, each counting photons (or neutrinos) within a small frequency window. In this way, one can study the consequences of the frequency-dependence of the opacities. Here, we will show that, even with this improvement, M1 closure still possesses vanishing shear viscosity.

For clarity, we will be focusing on the (Newtonian\footnote{Actually, the approach of \cite{Vartanyan:2018iah} is semi-relativistic, because it accounts for gravitational redshift. However, here we are working in flat spacetime, so the theory is effectively Newtonian, although, of course, radiation is never truly Newtonian.}) radiation-hydrodynamic description used in supernova simulations by \citet{Vartanyan:2018iah}. To ease the comparison, we will adopt exactly the same notation as in Appendix A of \cite{Vartanyan:2018iah}. The degrees of freedom of the theory are
$\Psi=\{\rho,v^j,e,Y_e,E_{s\varepsilon},F_{s\varepsilon}^j\}$, representing respectively the mass density, flow velocity, and energy density of matter, the electron fraction, and the energy density and flux per unit frequency $\varepsilon/(2\pi)$ of the neutrino species $s$. The equations of motion of the system are equations (A1)-(A6) of \cite{Vartanyan:2018iah}, namely
\newpage
\begin{equation}\label{burrows1}
\begin{split}
\partial_t \rho +\partial_k (\rho v^k)={}& 0 \, ,\\
\partial_t(\rho v_j)+\partial_k (\rho v^k v_j +P \delta^k_j)={}& \sum_s \int_0^\infty (\kappa_{s\varepsilon}{+}\sigma^{\text{tr}}_{s\varepsilon})F_{s\varepsilon j } d\varepsilon \, , \\
\partial_t \bigg[\rho\bigg( e + \dfrac{v^l v_l}{2}\bigg) \bigg] +  \partial_k \bigg[\rho v^k \bigg( e+\dfrac{v^l v_l}{2} +\dfrac{P}{\rho}  \bigg) \bigg] ={}& -\sum_s \int_0^\infty \bigg( j_{s\varepsilon}-\kappa_{s\varepsilon}E_{s\varepsilon}-v^k (\kappa_{s\varepsilon}{+}\sigma^{\text{tr}}_{s\varepsilon})F_{s\varepsilon k } \bigg) d\varepsilon  \, , \\
\partial_t(\rho Y_e)+ \partial_k(\rho Y_e v^k)={}& \sum_s \int_0^\infty \xi_{s\varepsilon} (j_{s\varepsilon}-\kappa_{s\varepsilon}E_{s\varepsilon})d\varepsilon \, , \\
\partial_t E_{s\varepsilon} +\partial_k(F_{s\varepsilon}^k +v^k E_{s\varepsilon})-\partial_k v^j \dfrac{\partial}{\partial \ln \varepsilon} P^k_{s\varepsilon j}={}& j_{s\varepsilon}-\kappa_{s\varepsilon}E_{s\varepsilon} \, , \\
\partial_t F_{s\varepsilon j}+ \partial_k (P^k_{s\varepsilon j}+v^kF_{s\varepsilon j})+\partial_j v^k F_{s\varepsilon k}-\partial_l v^k \dfrac{\partial}{\partial\varepsilon}(\varepsilon Q^l_{s\varepsilon jk})={}& -(\kappa_{s\varepsilon}{+}\sigma^{\text{tr}}_{s\varepsilon})F_{s\varepsilon j } \, , \\ 
\end{split}
\end{equation}
where $P$ is the matter's pressure, while $\{\kappa_{s\varepsilon},\sigma^{\text{tr}}_{s\varepsilon},j_{s\varepsilon},\xi_{s\varepsilon},P^k_{s\varepsilon j},Q^l_{s\varepsilon jk} \}$ are respectively the absorption opacity, the scattering opacity, the emission coefficient, the electron-lepton number, the pressure tensor and the heat tensor of neutrinos of type $s$ with energy $\varepsilon$. More details can be found in \cite{Vartanyan:2018iah,Vaytet2011}.

Let us show that the stationary state
\begin{equation}\label{burrows2}
\begin{cases}
v^j=\big(0,0,v(x^1)\big) \spc (\text{where }v\text{ is an arbitrary function})\\
F^j_{s\varepsilon}=0\\
(\rho,e,Y_e,E_{s\varepsilon})=\text{const}=\text{``local thermodynamic equilibrium''}
\end{cases}
\end{equation}
is a solution of the equations of motion. Here, ``local thermodynamic equilibrium'' means that $E_{s\varepsilon}=E_{s\varepsilon}^{LTE}(\rho,e,Y_e)$ is a Fermi-Dirac distribution, whose chemical potential is such that all reactions are in equilibrium at each event. 

To show that \eqref{burrows2} solves \eqref{burrows1}, we just need to plug the former in the latter. Let us first notice that, since in \eqref{burrows2} matter is in local thermodynamic equilibrium, the Kirchhoff-Planck relation holds, namely $j_{s\varepsilon}=\kappa_{s\varepsilon}E_{s\varepsilon}^{LTE}$ \cite{mihalas_book}. This, together with the fact that $F^j_{s\varepsilon}$ vanishes, implies that all right-hand sides of \eqref{burrows1} vanish along \eqref{burrows2}. Furthermore, considering that \eqref{burrows2} does not depend on time, we can drop all time derivatives. Therefore, the system \eqref{burrows1} becomes
\begin{equation}\label{burrows3}
\begin{split}
\partial_k (\rho v^k)={}& 0 \, ,\\
\partial_k (\rho v^k v_j +P \delta^k_j)={}&0 \, , \\
\partial_k \bigg[\rho v^k \bigg( e+\dfrac{v^l v_l}{2} +\dfrac{P}{\rho}  \bigg) \bigg] ={}& 0 \, , \\
\partial_k(\rho Y_e v^k)={}& 0 \, , \\
\partial_k(v^k E_{s\varepsilon})-\partial_k v^j \dfrac{\partial}{\partial \ln \varepsilon} P^k_{s\varepsilon j}={}& 0 \, , \\
\partial_k P^k_{s\varepsilon j}-\partial_l v^k \dfrac{\partial}{\partial\varepsilon}(\varepsilon Q^l_{s\varepsilon jk})={}& 0 \, . \\ 
\end{split}
\end{equation}
Now, the first four equations are trivially satisfied, because the only non-vanishing derivative is $\partial_1$ and the only non-vanishing component of the velocity is $v^3$. Hence, we are only left with the two following equations:
\begin{equation}\label{upsuzzuz}
\begin{split}
\partial_1 v^3 \dfrac{\partial}{\partial \ln \varepsilon} P^1_{s\varepsilon 3}={}& 0 \, , \\
\partial_1 P^1_{s\varepsilon j}-\partial_1 v^3 \dfrac{\partial}{\partial\varepsilon}(\varepsilon Q^1_{s\varepsilon j3})={}& 0 \, . \\ 
\end{split}
\end{equation}
Now we invoke the closure, according to which $P^k_{s\varepsilon j}=E_{s\varepsilon }\delta^j_k/3$ and $Q^l_{s\varepsilon jk}=0$ whenever $F^j_{s\varepsilon}=0$ \cite{Vaytet2011}. Then, all the three terms in equations \eqref{upsuzzuz} vanish, meaning that the system \eqref{burrows1} is indeed fulfilled. This completes our proof: The system \eqref{burrows1} admits, as a solution, a stationary shear wave, which survives forever and never decays.

\newpage
\section{Grey dynamics}

\subsection{Derivation and limiting regimes}
\begin{figure}
\begin{center}
\includegraphics[width=0.50\textwidth]{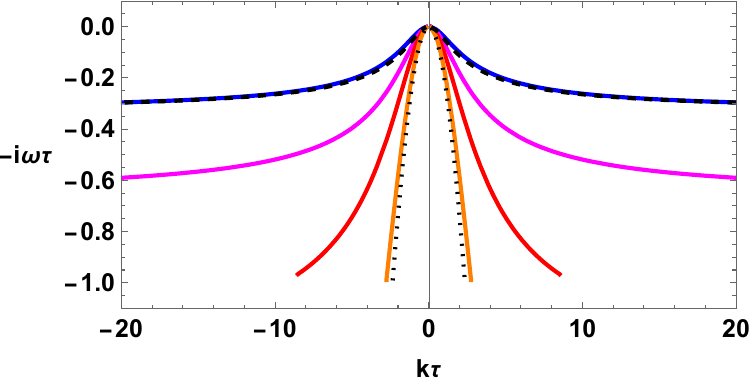}
\caption{Dispersion relation $\omega(k)$ of a sinusoidal shear wave in a grey fluid, with radiation-driven viscosity. Continuous lines: Exact formula \eqref{labireora}, for $\lambda{=}0.5$ (blue), $\lambda{=}1$ (magenta), $\lambda{=}2$ (red), and $\lambda{=}10$ (orange). Dashed line: Approximate formula \eqref{greyass} for matter-dominated systems, with $\lambda{=}0.5$. Dotted line: Radiation-dominated formula \eqref{rta}, recovered from \eqref{labireora} as $\lambda {\rightarrow} \infty$.}
	\label{fig:approximation}
	\end{center}
\end{figure}

Let us consider again the implicit function
\begin{equation}
\omega +\dfrac{i}{W} \sum_n \int_0^{\infty}  \bigg[\dfrac{2}{3}+\dfrac{1{-}i\omega \tau_n}{(k\tau_n)^2}-\bigg(1+ \dfrac{(1{-}i\omega \tau_n)^2}{(k\tau_n)^2} \bigg) \dfrac{1}{k\tau_n} \arctan\bigg( \dfrac{k\tau_n}{1{-}i\omega \tau_n}\bigg)\bigg] \rho_n dE=0 \, .
\end{equation}
This equation simplifies considerably if we assume grey interactions, i.e. $\tau_n(E)=\text{const}\equiv \tau \in \mathbb{R}$. Under this assumption, defined the dimensionless parameter 
\begin{equation}
\lambda= \dfrac{\tau}{W} \sum_n \int_0^{\infty}  \rho_n(E) dE \equiv \dfrac{3\tau }{2W} \eta^{(-1)} \, ,
\end{equation}
we have
\begin{equation}\label{gryuzco}
\omega \tau + i \lambda  \bigg[\dfrac{2}{3}+\dfrac{1{-}i\omega \tau}{(k\tau)^2}-\bigg(1+ \dfrac{(1{-}i\omega \tau)^2}{(k\tau)^2} \bigg) \dfrac{1}{k\tau} \arctan\bigg( \dfrac{k\tau}{1{-}i\omega \tau}\bigg)\bigg]=0 \, .
\end{equation}
This implicit function admits an analytical parametric solution $\{k(r),\omega(r)\}:\mathbb{R}\rightarrow \mathbb{R} \times i\mathbb{R}$, with
\begin{equation}\label{labireora}
\begin{cases}
 \omega(r)= -\dfrac{i}{\tau} \bigg[\dfrac{1}{2}{+}\dfrac{\lambda}{3}-\sqrt{\bigg( \dfrac{1}{2}{-}\dfrac{\lambda}{3} \bigg)^2 -\lambda \bigg[\dfrac{1}{r^2}{-}\bigg( 1{+}\dfrac{1}{r^2}  \bigg)\dfrac{\arctan(r)}{r} \bigg] }  \, \, \bigg]     \, , \\
 k(r)= +\dfrac{r}{\tau} \bigg[\dfrac{1}{2}{-}\dfrac{\lambda}{3}+\sqrt{\bigg( \dfrac{1}{2}{-}\dfrac{\lambda}{3} \bigg)^2 -\lambda \bigg[\dfrac{1}{r^2}{-}\bigg( 1{+}\dfrac{1}{r^2}  \bigg)\dfrac{\arctan(r)}{r} \bigg] }  \, \, \bigg] \, , \\
\end{cases}
\end{equation}
where it can be easily verified that $r=k\tau/(1-i\omega\tau)$.
Using dimensional analysis, one finds that $\lambda$ scales like $T^{00}_{\text{rad}}/T^{00}_\text{fluid}$. For example, in the case of photon radiation subject to absorption and isoenergetic scattering, we have
\begin{equation}
\lambda=\dfrac{2 \times \text{``radiation energy density''}}{\text{``fluid's relativistic enthalpy density''}}  \, .  
\end{equation}
Hence, in matter-dominated grey systems, one has that $\lambda \ll 1$, and equation \eqref{labireora} reduces to
\begin{equation}\label{greyass}
\omega(k)  = -i \dfrac{\lambda}{\tau}  \bigg[\dfrac{2}{3}+\dfrac{1}{(k\tau)^2}-\bigg(1+ \dfrac{1}{(k\tau)^2} \bigg) \dfrac{\arctan (k\tau)}{k\tau}\bigg] \, ,
\end{equation}
while in radiation-dominated grey systems one has that $\lambda \gg 1$, and \eqref{labireora} reduces to
\begin{equation}\label{rta}
\begin{cases}
 \omega(r)= -\dfrac{3i}{2\tau} \bigg[ \dfrac{2}{3} +\dfrac{1}{r^2}-\bigg(1+\dfrac{1}{r^2} \bigg) \dfrac{\arctan(r)}{r} \bigg]     \, , \\
 k(r)= +\dfrac{3}{2\tau r} \bigg[ \bigg(r+\dfrac{1}{r}  \bigg)\arctan(r)-1 \bigg] \, . \\
\end{cases}
\end{equation}
All dispersion relations \eqref{labireora}, \eqref{greyass}, and \eqref{rta} are plotted in figure \ref{fig:approximation}. The qualitative transition from matter-dominated systems to radiation-dominated systems occurs at $\lambda=3/2$. In fact, below $3/2$, the function $k(r)$ is unbounded, meaning that the model is ``hydrodynamic'' at all lengthscales. This is an indication that the motion is dominated by the fluid component, which evolves hydrodynamically for all $k$. Instead, above $3/2$, the function $k(r)$ is bounded, meaning that hydrodynamics breaks down at the cutoff scale
\begin{equation}
L_c =k(r{=}\infty)^{-1}=\dfrac{2\tau}{\pi} \bigg(\dfrac{2}{3}-\dfrac{1}{\lambda} \bigg) \, .
\end{equation}
This is a sign that radiation (which is governed by kinetic theory) dominates the motion, and the system transitions to a ``non-hydrodynamic'' regime below $L_c$.

\newpage

\subsection{The radiation-dominated regime}

Let us consider the radiation-dominated case (i.e. $\lambda =\infty$), where the curve $\{k(r),\omega(r)\}$ is given by system \eqref{rta}. The two equations in \eqref{rta} can be combined to give the following implicit function:
\begin{equation}
\dfrac{2}{3}+\dfrac{1{-}i\omega \tau}{(k\tau)^2}-\bigg(1+ \dfrac{(1{-}i\omega \tau)^2}{(k\tau)^2} \bigg) \dfrac{1}{k\tau} \arctan\bigg( \dfrac{k\tau}{1{-}i\omega \tau}\bigg) =0 \, ,
\end{equation}
which could also be obtained directly from \eqref{gryuzco}. Let us define $y(k)=1-i\tau \omega(k)$. Then, the above equation may be equivalently rewritten as follows:
\begin{equation}\label{kty}
\dfrac{k\tau}{y} = \tan \bigg[ \dfrac{2(k\tau)^3 +3 k\tau y}{3(k^2\tau^2+y^2 )} \bigg]  \, .  
\end{equation}
Now, recalling the trigonometric identity
\begin{equation}
\dfrac{2\tan (\theta)}{1-\tan^2 (\theta)} =\tan(2\theta)  \, ,  
\end{equation}
we can rearrange \eqref{kty} in the following form:
\begin{equation}
\dfrac{2k\tau y}{y^2-k^2 \tau^2} = \tan \bigg[ \dfrac{4(k\tau)^3 +6 k\tau y}{3(k^2\tau^2+y^2 )} \bigg] \, .   
\end{equation}
This is precisely the dispersion relation of shear waves in ultrarelativistic kinetic theory in the relaxation-type approximation (RTA) \cite[\S III.B]{Romatschke:2015gic}. The physical interpretation is clear: At large $\lambda$, the energy-momentum tensor is dominated by the radiation gas, which is governed by kinetic theory. The fluid component only acts as a source of thermalization, through a collision integral that is mathematically identical (in incompressible grey systems) to the RTA.

\newpage
\section{Decay of a vortex by radiative friction}

In this section, we give an example of a two-dimensional solenoidal (i.e. incompressible) flow that decays due to radiative friction. This can be expressed as a Fourier superposition of transversal plane waves, meaning that our results still apply (with the same dispersion relations), and we can use our model to describe the full evolution. 

\subsection{Initial state and Navier-Stokes solution}

\begin{figure}[b!]
    \centering
    \includegraphics[width=0.48\linewidth]{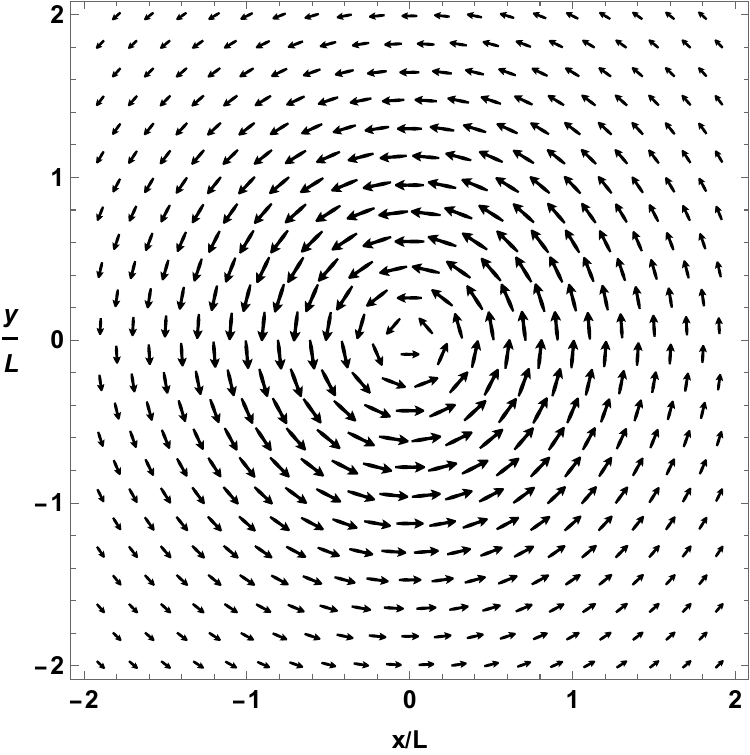}
    \includegraphics[width=0.49\linewidth]{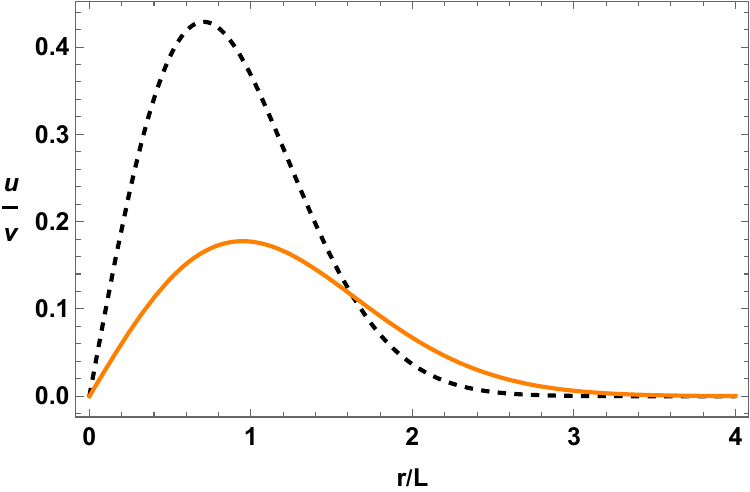}
    \caption{Let panel: Initial state of our vortex, as given by equation \eqref{iniziovortex}. The motion is perfectly incompressible and circular. The tangential speed is maximal at $r=L/\sqrt{2}$, where $r=\sqrt{x^2{+}y^2}$ is the distance from the axis of rotation. Right panel: Tangential speed as a function of $r$ along the Navier-Stokes solution \eqref{navotex}. The dashed line is the initial profile. The orange line is the shape at $\nu t=0.3$. The location of the maximum moves outwards, being located at $r=\sqrt{(L^2{+}4\nu t)/2}$.}
    \label{fig:NavierVoertex}
\end{figure}

We consider the vector potential
\begin{equation}
\mathbb{A}= \dfrac{vL}{2} e^{-(x^2+y^2)/L^2}
\begin{pmatrix}
0 \\
0 \\
1 \\
\end{pmatrix}
\end{equation}
where $v$ and $L$ are some positive constants. This generates the incompressible flow
\begin{equation}\label{iniziovortex}
\textbf{u}=\nabla \times \mathbb{A}=\dfrac{v}{L} e^{-\frac{x^2+y^2}{L^2}} 
\begin{pmatrix}
-y \\
x\\
0
\end{pmatrix}\, ,
\end{equation}
which has the shape of a localized cylindrically symmetric vortex (see figure \ref{fig:NavierVoertex}, left panel). Its radius is approximately $L$, which defines the hydrodynamic lengthscale. Taking this shape as our initial condition, we can solve the linearised incompressible Navier-Stokes equation analytically, and we obtain (defining the kinematic viscosity $\nu =\eta^{(1)}/W$)
\begin{equation}\label{navotex}
\textbf{u}(t,x,y)=vL^3 \, \dfrac{e^{-\frac{x^2+y^2}{L^2+4\nu t}}}{(L^2+4\nu t)^2}
\begin{pmatrix}
-y \\
x\\
0
\end{pmatrix}\, .
\end{equation}
As can be seen from figure \ref{fig:NavierVoertex} (right panel), the vortex decays by diffusing its angular momentum to the outer layers.

\subsection{Non-Newtonian corrections}

\begin{figure}[b!]
    \centering
\includegraphics[width=0.45\linewidth]{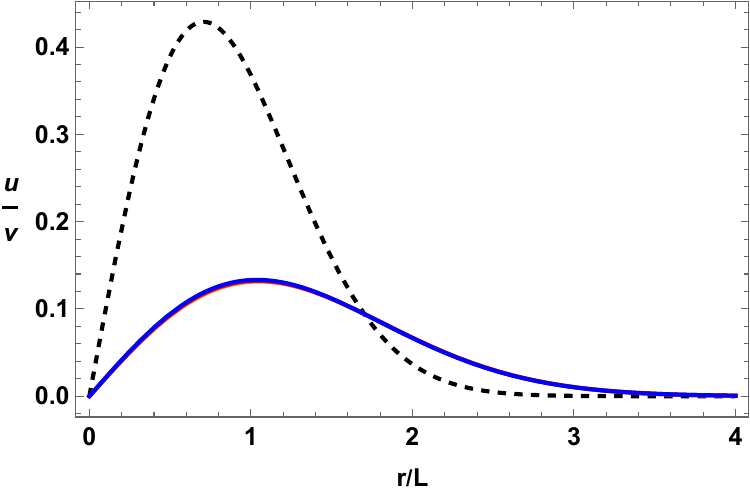}
\includegraphics[width=0.45\linewidth]{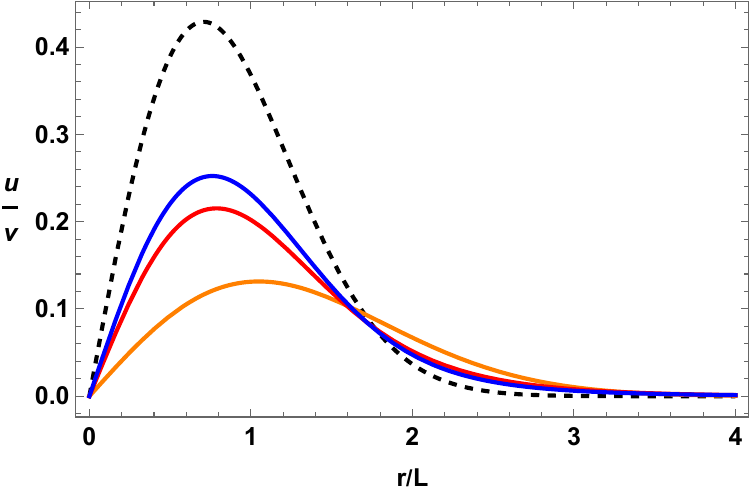}
\includegraphics[width=0.45\linewidth]{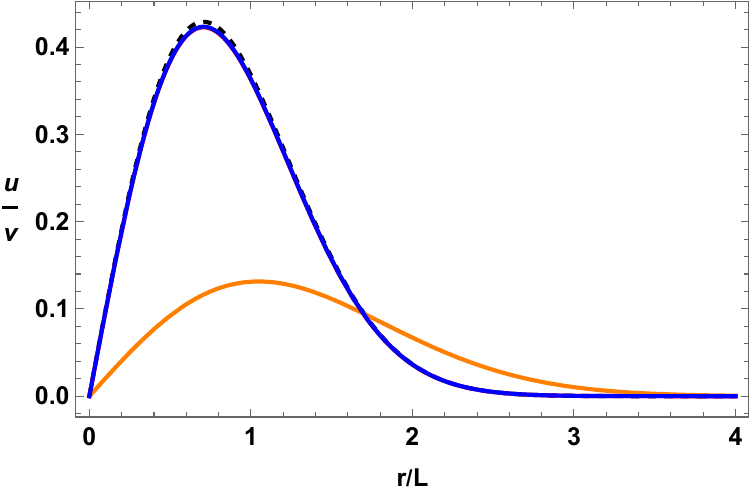}
\includegraphics[width=0.45\linewidth]{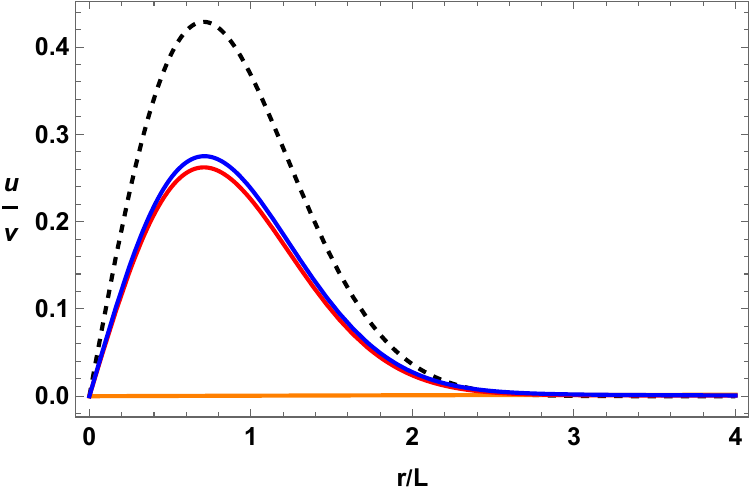}
    \caption{Profile of the tangential speed of the vortex according to \eqref{utr} for different choices of $\{\tau/L,\nu t/L^2 \}$, respectively up-left:$\{0.1,0.3 \}$, up-right:$\{1,0.3 \}$, down-left:$\{10,0.3 \}$, down-right $\{10,10\}$. Each plot shows the initial state (dashed), the solution of the Boltzmann equation (blue), the non-Newtonian model provided in the main text (red), and the Navier-Stokes solution (orange). The quantity $\tau/L$ is the radiative Knudsen number, and it quantifies the size of the non-Newtonian corrections. For small $\tau/L$, all curves overlap. For $\tau/L=1$, the non-Newtonian model fares better than Navier-Stokes, but it is not perfect. At large $\tau/L$, Navier-Stokes dramatically overestimates the decay, while the non-Newtonian model is again very accurate.}
    \label{fig:nonewtvortex}
\end{figure}

\vspace{-0.2cm}
Let us now compute the non-Newtonian corrections due to the finite photon mean free path. To this end, we first note that the initial shape \eqref{iniziovortex} can be expressed as the following Fourier integral: 
\vspace{-0.2cm}
\begin{equation}
\mathbf{u}(0,x,y) = \dfrac{iv L^3}{8\pi} \int dk_x dk_y \, e^{-\frac{L^2}{4}(k_x^2 +k_y^2)+i(k_x x+k_y y)}
\begin{pmatrix}
k_y \\
-k_x \\
0 \\
\end{pmatrix}\, .
\end{equation}
Each plane wave contributing to this integral is transversal, namely $\mathbf{k}\cdot \mathbf{u}(\mathbf{k})=0$, which follows from the vector field being solenoidal. Hence, all these waves evolve according to the equations shown in the main text, up to a rotation in space. In particular, they all decay with the dispersion relation $\omega(\sqrt{k_x^2 +k_y^2})$ computed in the main text, giving
\vspace{-0.2cm}
\begin{equation}
\mathbf{u}(t,x,y) = \dfrac{iv L^3}{8\pi} \int dk_x dk_y \, e^{-\frac{L^2}{4}(k_x^2 +k_y^2)+i(k_x x+k_y y)-i\omega(\sqrt{k_x^2+k_y^2})t}
\begin{pmatrix}
k_y \\
-k_x \\
0 \\
\end{pmatrix}\, .
\end{equation}
To determine the tangential speed of the vortex as a function of the distance, we can just set $y=0$ and $x=r$. Then, $ u^x$ vanishes, and we have that $ u=| \mathbf{u}|= u^y$. Working in polar coordinates, $\{k_x=k\cos \theta,k_y=k\sin \theta\}$, we can integrate out the angle, and we finally obtain the expression below, which can be used to compare different approaches (see figure \ref{fig:nonewtvortex}):
\vspace{-0.2cm}
\begin{equation}\label{utr}
u(t,r) = \dfrac{v L^3}{4} \int_0^{+\infty} k^2 J_1(kr)  e^{-\frac{L^2k^2}{4}-i\omega(k)t} dk 
\, .
\end{equation}

\label{lastpage}

\end{document}